\documentclass[reqno,12pt]{amsart}

\usepackage{amsmath}
\usepackage{latexsym}
\usepackage{amssymb}
\usepackage{hyperref}
\usepackage{graphicx}
\usepackage{epstopdf}
\usepackage{ifpdf}
\ifpdf
\fi

\makeatletter
 
 \newcommand{\Rmnum}[1]{\expandafter\@slowromancap\romannumeral #1@}
 \makeatother

\newtheorem{theorem}{Theorem}[section]

\newtheorem{proposition}[theorem]{Proposition}

\newtheorem{remark}[theorem]{Remark}

\newenvironment{sequation}{\begin{equation}\small}{\end{equation}}

\newcommand{\R}{{\mathbb R}}

\newcommand{\Z}{{\mathbb Z}}
\newcommand{\C}{{\mathbb C}}

\newcommand{\nn}{\nonumber}
\newcommand{\be}{\begin{equation}}
\newcommand{\ee}{\end{equation}}
\newcommand{\bea}{\begin{eqnarray}}
\newcommand{\eea}{\end{eqnarray}}
\newcommand{\ba}{\begin{array}}
\newcommand{\ea}{\end{array}}

\newcommand{\id}{\mathbb{I}}

\newcommand{\re}{\mathrm{Re}}
\newcommand{\im}{\mathrm{Im}}

\newcommand{\cF}{\mathcal{F}}

\newcommand{\vphi}{\varphi}
\newcommand{\sig}{\sigma}
\newcommand{\Sig}{\Sigma}
\newcommand{\lam}{\lambda}

\newcommand{\gam}{\gamma}
\newcommand{\Gam}{\Gamma}
\newcommand{\om}{\omega}
\newcommand{\Om}{\Omega}
\newcommand{\x}{\xi}
\newcommand{\dta}{\delta}
\newcommand{\Dta}{\Delta}
\newcommand{\al}{\alpha}
\newcommand{\tha}{\theta}

\newcommand{\varplon}{\varepsilon}

\numberwithin{equation}{section}

\begin{document}

\title[GI-DNLS Long-time asymptotic with step-like initial value]{Long-time asymptotic for the derivative nonlinear Schr\"odinger equation with step-like initial value}

\author[J.Xu]{Jian Xu}
\address{School of Mathematical Sciences\\
Fudan University\\
Shanghai 200433\\
People's  Republic of China}
\email{11110180024@fudan.edu.cn}

\author[E.Fan]{Engui Fan*}
\address{School of Mathematical Sciences, Institute of Mathematics and Key Laboratory of Mathematics for Nonlinear Science\\
Fudan University\\
Shanghai 200433\\
People's  Republic of China}
\email{correspondence author: faneg@fudan.edu.cn}

\author[J.Xu]{Yong Chen}
\address {Shanghai Key Laboratory of Trustworthy Computing\\
 East China Normal University,\\
Shanghai 200062, People¡¯s Republic
of China.}
\email{ychen@sei.ecnu.edu.cn}


\date{\today}

\begin{abstract}

We consider the Cauchy problem for the Gerdjikov-Ivanov(GI) type of the derivative nonlinear Schr\"odinger (DNLS) equation: $$iq_t+q_{xx}-iq^2\bar{q}_x+\frac{1}{2}|q|^4{q}=0.$$ with steplike initial data: $q(x,0)=0$ for $x\le 0$ and $q(x,0)=Ae^{-2iBx}$ for $x>0$,where $A>0$ and $B\in \R$ are constants.The paper aims at studying the long-time asymptotics of the solution to this problem.We show that there are four regions in the half-plane $-\infty<x<\infty,t>0$,where the asymptotics has qualitatively different forms:a slowly decaying self-similar wave of Zakharov-Manakov type for $x>-4tB$, a plane wave region:$x<-4t(B+\sqrt{2A^2(B+\frac{A^2}{4})})$, an elliptic region:$-4t(B+\sqrt{2A^2(B+\frac{A^2}{4})})<x<-4tB$. The main tool is the asymptotic analysis of an associated matrix Riemann-Hilbert problem.

\end{abstract}

\maketitle

\section{Introduction}

The classical, mathematical model for non-linear pulse propagation in the picosecond time scale in the anomalous dispersion regime in an isotropic, homogeneous, lossless, non-amplifying, polarization-preserving single-mode optical fibre is the non-linear
Schr\"odinger(NLS) equation \cite{ft}. However, in the subpicosecond-femtosecond time scale, experiments and theories on the propagation of high-power ultrashort pulses in long monomode optical fibres have shown that the NLS equation is no longer valid and that additional non-linear terms (dispersive and dissipative) and higher-order linear dispersion should be taken into account, you can see \cite{avkahv3} and the references therein. In this case, subpicosecond-femtosecond pulse propagation is described (in dimensionless and normalized form) by the following non-linear evolution equation (NLEE)
\be \label{NLEE}
iu_\x+\frac{1}{2}u_{\tau \tau}+|u|^2u+is(|u|^2u)_\tau=-i\tilde \Gam u+i\tilde \dta u_{\tau \tau \tau}+\frac{\tau_n}{\tau_0}u(|u|^2)_\tau,
\ee
where $u$ is the slowly varying amplitude of the complex field envelope, $\x$ is the propagation distance along the fibre length, $\tau$ is the time measured in a frame of reference moving with the pulse at the group velocity (the retarded frame), $s (> 0)$ governs the effects due to the intensity dependence of the group velocity (self-steepening), $\tilde \Gam$
is the intrinsic fibre loss, $\tilde \dta$
governs the effects of the third-order linear dispersion, and $\frac{\tau_n}{\tau_0}$, where $\tau_0$ is the normalized
input pulsewidth and $\tau_n$ is related to the slope of the Raman gain curve (assumed to vary
linearly in the vicinity of the mean carrier frequency, $\om_0$), governs the soliton self-frequency
shift (SSFS) effect, \cite{avkahv3} and the references therein.
\par
We set the right-hand side of (\ref{NLEE}) equal to zero, we obtain the following equation,
\be \label{NLEEzero}
iu_\x+\frac{1}{2}u_{\tau \tau}+|u|^2u+is(|u|^2u)_\tau=0,
\ee
This equation is related to the Kaup-Newell type of derivative nonlinear Schr\"odinger equation,
\be \label{eq:KNDNLS+}
iq_t(x,t)=-q_{xx}(x,t)+(\bar q q^2)_x
\ee
by change of variables
\[
u(\x,\tau)=q(x,t)e^{i(\frac{t}{4s^4}-\frac{x}{2s^2})},\quad \x=\frac{t}{2s^2},\quad \tau=-\frac{x}{2s}+\frac{t}{2s^3}.
\]
And we note that if we replace $x$ by $-x$,equation (\ref{eq:KNDNLS+}) changes into
\be \label{eq:KNDNLS-}
iq_t(x,t)=-q_{xx}(x,t)-(\bar q q^2)_x.
\ee
But, we also know if we formulate a Riemann-Hilbert problem for the solution of the inverse spectral problem of the equation (\ref{eq:KNDNLS-}), we find we cannot find solutions of its spectral problem which approach the $2\times 2$ identity matrix $\id$ as $k\rightarrow \infty$.It is well-known that  there are  three kinds of celebrated DNLS equations,  including   Kaup-Newell equation ( i.e Eq.(\ref{eq:KNDNLS-})),
Chen-Lee-Liu equation \cite{chen2}
$$iq_t+q_{xx}+i|q|^2q_x=0,$$
and Gerdjikov-Ivanov(GI) equation \cite {kundu,gi}
\be \label{eq:GIDNLS}
iq_t+q_{xx}-iq^2\bar{q}_x+\frac{1}{2}|q|^4{q}=0
\ee
It has been  found that they may be transformed into each
other by  gauge transformations \cite{kundu, fan2}. And in \cite{gi}, the GI-type has the required property of the solutions of its spectral problem which approach the $2\times 2$ identity matrix $\id$ as $k\rightarrow \infty$. So,we focus on the GI-type of derivative nonlinear Schr\"odinger equation. In the following of this paper we also name the GI-type DNLS equation as DNLS equation.
\par
Initial value problems for nonlinear evolution equations with step-like initial data have attracted much attention since the early 1970s \cite{gp,kh,kk,v}, but only a few rigorous results concerning the long-time behavior of solutions of such problems were available.In 1980s-1990s, a considerable progress was achieved following the development of the theory of Whitham deformations \cite{w} and the analysis of matrix Riemann-Hilbert problem representations of solutions of initial value problems, see \cite{b1,b2,n} and references therein.Most complete results,obtained by using this approach,were related to integrable equations,for which linear operators from the associated Lax pair were self-adjoint and thus their spectrum was real.In \cite{b2},Bikbaev considered the case of the focusing nonlinear Schr\"odinger equation,which required the development of a much more complicated complex form of the theory of Whitham deformations.
\par
A completely rigorous approach for studying asymptotics of solutions of integrable nonlinear equations was introduced by Deift and Zhou \cite{dz}(this approach was inspired by earlier works of Manakov \cite{m} and Its \cite{i};see \cite{diz} for a detailed historical review) and further extended by Deift,Venakides,and Zhou \cite{dvz1,dvz2}. This approach is based on the development of the nonlinear steepest descent method for Riemann-Hilbert problems associated with integrable nonlinear equations. Being originally introduced for studying initial value problems with decaying initial data, this approach was recently adapted by Buckingham and Venakides \cite{bv} to problems with shock-type oscillating initial data for focusing nonlinear Schr\"odinger equation. A central role in this development is played by the so-called $g-$function mechanism allowing to deform the original Riemann-Hilbert problem to a form that can be asymptotically treated with the help of associated singular integral equations.
\par
The Riemann-Hilbert problem approach to initial value problems with nondecaying step-like initial data shares many issues with the adaptation of this approach for studying initial-boundary value problems with non-decaying boundary data \cite{abmik,abmk,abmks}.However,there is an important difference: in the latter case,the construction of the associated Riemann-Hilbert problem normally requires the knowledge of spectral functions associated with overspecified initial and boundary data,which leads to the fact that results(in particular,the asymptotic results,see \cite{abmik}) have,in a certain sense,a conditional character.As for the initial value problems of the type considered in this paper,the Riemann-Hilbert construction requires only initial data,and thus,the issue of overdetermination does not arise.
\par
In this paper,we consider a pure step-like initial value problem for the DNLS equation:
\begin{subequations} \label{DNLSandInit}
\be \label{eq:DNLS}
iq_t+q_{xx}-iq^2\bar{q}_x+\frac{1}{2}|q|^4{q}=0,\qquad x\in \R,t>0,
\ee
\be \label{eq:initial}
q(x,0)=q_0(x)=\left\{\ba{lr}0&\mbox{if }x\geq 0,\\
                            Ae^{-2iBx}&\mbox{if }x<0,
                            \ea \right.
\ee
\end{subequations}
where $A>0$ and $B\in \R$ are some constants. Kitaev and Vartanian got the leading order long-time asymptotic for the KN-type of DNLS equation with the decaying initial value,in \cite{avkahv}, and the higher order long-time asymptotic in \cite{avkahv3}.
\par
Since the DNLS equation (\ref{eq:DNLS}) has a plane wave solution
\be \label{psolu}
q^p(x,t)=Ae^{-2iBx+2i\om t},
\ee
with
\be \label{om}
\om:=A^2B-2B^2+\frac{A^4}{4},
\ee
which is consistent with (\ref{eq:initial}) for $x<0$,that is,$q^p(x,0)=q_0(x)$,we assume that the solution $q(x,t)$ of the initial value problem (\ref{eq:DNLS}) evaluated at any $t>0$ has the following behavior as $x\rightarrow \pm \infty$:
\be \label{qasymp+}
q(x,t)=o(1),\qquad x\rightarrow +\infty,
\ee
\be \label{qasymp-}
q(x,t)=q^p(x,t)+o(1),\qquad x\rightarrow -\infty,
\ee
where $o(1)$ means sufficiently fast decay to $0$.This assumption can be justified a posteriori,by evaluating the large-$x$ behavior of the solution of the Riemann-Hilbert problem formulated in Section 3.
\par
Recently, in \cite{abmks1},A.Boutet de Monvel,V.P.Kotlyarov, and D.Shepelsky considered the long-time dynamics of the initial value problem for the focusing nonlinear Schr\"odinger equation with step-like data.The strategy of the Riemann-Hilbert problem deformations that we adopt in this paper is similar,though not identical,to that in \cite{bv}.In particular,the realization of the $g-$function mechanism is different as well as the resulting asymptotic picture.
\par
As we have already mentioned, the main tool available now for studying rigorously the long-time asymptoitcs of solutions of initial and initial boundary value problems for integrable nonlinear equations is the asymptotic analysis of associated Riemann-Hilbert problems,whose construction involves dedicated solutions of the system of two linear equations,the Lax pair associated with the integrable nonlinear equation.
\par
For the DNLS equation (\ref{eq:DNLS}), a Lax pair is as follows \cite{avkahv}:
\be \label{laxpair}
\begin{split}
&\Psi_x(x,t;k)=M(x,t;k)\Psi(x,t;k),\\
&\Psi_t(x,t;k)=N(x,t;k)\Psi(x,t;k),
\end{split}
\ee
where
\be \label{MN}
\begin{split}
&M(x,t;k)=-ik^2\sig_3+kQ+\frac{i}{2}|q|^2\sig_3,\\
&N(x,t;k)=-2ik^4\sig_3+2k^3Q+ik^2|q|^2\sig_3-ikQ_x\sig_3+\frac{i}{4}|q|^4\sig_3+\frac{1}{2}(q{\bar q}_x-\bar qq_x)\sig_3,
\end{split}
\ee
with $\sig_3=\left(\ba{lc}1&0\\0&-1\ea \right)$, and $\Psi(x,t;k)$ is a $2 \times 2$ matrix-value function,$k\in \C$ is a spectral parameter, and the matrix coefficient $Q$ is expressed in terms of a scalar function $q$:
\be \label{Q}
Q=\left( \ba{lc}0&q\\-\bar q&0 \ea \right),
\ee
It is well-known \cite{avkahv} that this over-determined system of equations (\ref{laxpair}) is compatible if and only if $q(x,t)$ solves the DNLS equation (\ref{eq:DNLS}).
\par
In Section 2 we present these dedicated solutions(eigenfunctions) and associated spectral functions.All these functions are then used in Section 3 for constructing a basic Riemann-Hilbert problem,whose solution gives the solution of the initial value problem (\ref{eq:DNLS}),(\ref{eq:initial}).Section 4 develops the asymptotic analysis of this Riemann-Hilbert problem leading to asymptotic formulas for the solution of the original Cauchy problem (\ref{DNLSandInit}).

\section{Eigenfunctions}

Let $Q^p$ be defined by (\ref{Q}) with $q^p$ instead of $q$. A particular solution of the system (\ref{laxpair}),with $Q^p$ instead of $Q$,is given by
\be \label{Psip}
\Psi^p(x,t;k)=e^{i(\om t-Bx)\sig_3}E(k)e^{-i(xX(k)+t\Om(k))\sig_3},
\ee
where
\be \label{Xk}
X(k)=\sqrt{(k^2-B-\frac{A^2}{2})^2+k^2A^2},
\ee
\be \label{Omk}
\Om(k)=2(k^2+B)X(k).
\ee
\be \label{Ek}
E(k)=\frac{1}{2}\left(\ba{lc}\vphi(k)+\frac{1}{\vphi(k)}&\vphi(k)-\frac{1}{\vphi(k)}\\\vphi(k)-\frac{1}{\vphi(k)}&\vphi(k)+\frac{1}{\vphi(k)}\ea \right)
\ee
with
\be \label{vphi}
\vphi(k)=(\frac{k^2-B-\frac{A^2}{2}-ikA}{k^2-B-\frac{A^2}{2}+ikA})^{\frac{1}{4}},
\ee

The branch cut for $X$ and $\vphi$ is taken along the segment
\be \label{Xvphiseg}
\gam \cup \bar \gam:=\{k\in \C |k_1^2-k_2^2=B,k_1^2\le C^2 \},
\ee
where $\gam=\{k\in \C |k_1^2-k_2^2=B,k_1^2\le C^2,\im k^2>0\}$, $C^2=B+\frac{A^2}{4}$, $k_1=\re{k}$ and $k_2=\im{k}$. And the branches are fixed by the asymptotics:
\[
X(k)=k^2-B+O(\frac{1}{k^2}),\qquad \mbox{as }k\rightarrow \infty,
\]
\[
\vphi(k)=1+O(\frac{1}{k}),\qquad \mbox{as }k\rightarrow \infty.
\]
We find that $\Om(k)=2k^4+\om+O(\frac{1}{k}), \mbox{as }k\rightarrow \infty$. We also find that $\im{X(k)}=0$ is
\be \label{imXk}
k_1k_2(k_1^2-k_2^2-B)=0,
\ee
which is on
\be \label{imxk}
\Sig :=\R \cup i\R \cup \gam \cup \bar \gam.
\ee
Thus, for any $t\geq 0$, $\Psi^p(x,t;k)$ is bounded in $x$ if and only if $k\in \Sig$.
\par
Let $q(x,t)$ be a solution of the Cauchy problem  (\ref{eq:DNLS}),(\ref{eq:initial}) satisfying the asymptotic conditions (\ref{qasymp+}),(\ref{qasymp-}), and let $Q(x,t)$ and $Q^p(x,t)$ be defined by (\ref{Q}), in terms of $q$ and $q^p$, respectively.Define the $2\times 2$ matrix-value functions $\mu_j(x,t;k)$, $j=1,2$, $-\infty<x<\infty,0\le t<\infty$, as the solutions of the Volterra integral equations:
\be \label{mu1def}
\mu_1(x,t;k)=\id+\int_{+\infty}^x e^{ik^2(y-x)\sig_3}(kQ\mu_1)(y,t;k)e^{-ik^2(y-x)\sig_3},\qquad k^2\in \R,
\ee
\bea \label{mu2def}
\mu_2(x,t;k)&=&e^{i(\om t-Bx)\sig_3}E(k)  \\
&&+\int_{-\infty}^x \Gam^p(x,y,t,k)k[Q-Q^p](y,t)\mu_2(y,t,k)e^{-ik^2(y-x)\sig_3},k\in \Sig,\nn
\eea
where
\[
\Gam^p(x,y,t,k):=\Psi^p(x,t,k)[\Psi^p(y,t,k)]^{-1}.
\]
Note that $\Gam^p$ can be written in the form
\[
\Gam^p(x,y,t,k)=e^{i(\om t-Bx)\sig_3}G^p(x,y,k)e^{-i(\om t-By)\sig_3},
\]
where
\[
G^p(x,y,k)=\left(\ba{cc}\al+i(k^2-B-\frac{A^2}{2})\beta&-kA\beta \\kA\beta&\al-i(k^2-B-\frac{A^2}{2})\beta \ea \right),
\]
with
\[
\al=\cos[(y-x)X(k)],\qquad \beta=\frac{\sin[(y-x)X(k)]}{X(k)}.
\]
For any $(x,y)\in \R^2$,$G^p(x,y,k)$ is an entire function of $k$ with asymptotic behavior
\[
G^p(x,y,k)=e^{i(y-x)(k^2-B-\frac{A^2}{2})\sig_3}[\id+O(\frac{1}{k})],\qquad \mbox{as }k\rightarrow \infty,\quad \im{k^2}=0.
\]
The analytic properties of the $2\times 2$ matrices $\mu_j(x,t;k)$, $j=1,2$, that follow from (\ref{mu1def}) and (\ref{mu2def}) are collected in the following proposition.We denote by $\mu_j^{(1)}(x,t,k)$ and $\mu_j^{(2)}(x,t,k)$ the columns of $\mu_j(x,t;k)$.
\par
\begin{proposition}
The matrices $\mu_1(x,t;k)$ and $\mu_2(x,t;k)$ have the following properties:
\begin{enumerate}
\item $det \mu_1(x,t,k)=\mu_2(x,t;k)=1$.
\item The functions $\Phi(x,t,k)$ and $\Psi(x,t,k)$ defined by
      \[
      \Psi(x,t,k):=\mu_1(x,t,k)e^{-ik^2x\sig_3-2ik^4t\sig_3},
      \]
      \[
      \Phi(x,t,k):=\mu_2(x,t;k)e^{-ixX(k)\sig_3-it\Om(k)\sig_3}.
      \]
      satisfy the Lax pair equations (\ref{laxpair}).
\item $\mu_1^{(1)}(x,t,k)$ is analytic in $\im k^2<0$ and
      \[
      \mu_1^{(1)}(x,t,k)=\left(\ba{c}1\\0\ea \right)+O(\frac{1}{k}),\mbox{as }k\rightarrow \infty,\quad \im k^2 \le 0.
      \]
\item $\mu_1^{(2)}(x,t,k)$ is analytic in $\im k^2>0$ and
      \[
      \mu_1^{(2)}(x,t,k)=\left(\ba{c}0\\1\ea \right)+O(\frac{1}{k}),\mbox{as }k\rightarrow \infty,\quad \im k^2 \geq 0.
      \]
\item $\mu_2^{(1)}(x,t,k)$ is analytic in $\im k^2>0 \backslash \gam$,has a jump across $\gam$, and
      \[
      \mu_2^{(1)}(x,t,k)=\left(\ba{c}1\\0\ea \right)+O(\frac{1}{k}),\mbox{as }k\rightarrow \infty,\quad \im k^2 \geq 0.
      \]
\item $\mu_2^{(2)}(x,t,k)$ is analytic in $\im k^2<0 \backslash \bar \gam$,has a jump across $\bar \gam$, and
      \[
      \mu_2^{(2)}(x,t,k)=\left(\ba{c}0\\1\ea \right)+O(\frac{1}{k}),\mbox{as }k\rightarrow \infty,\quad \im k^2 \le 0.
      \]
\item Moreover,
      \[
      \mu_j^{(1)}(x,t,k)=\id+\frac{\tilde \mu(x,t)}{ik}+o(\frac{1}{k})
      \]
      as $k\rightarrow \infty$ along curves transversal to the real and image axis, where
      \[
      [\sig_3,\tilde \mu(x,t)]=\left(\ba{cc}0&q(x,t)\\-\bar q(x,t)&0\ea \right)
      \]
\item $\mu_2^{(2)}(x,t,k)(k-E)^{\frac{1}{4}}$ is boundary near $k=E$ and $\mu_2^{(2)}(x,t,k)(k-\bar E)^{\frac{1}{4}}$ is boundary near $k=\bar E$.
\end{enumerate}
\end{proposition}
Since the eigenfunctions $\Psi(x,t,k)$ and $\Phi(x,t,k)$ satisfy both equations of the Lax pair, we have
\be \label{psiphirel}
\Phi(x,t,k)=\Psi(x,t,k)S(k),\qquad k^2\in \R,
\ee
where $S(k)$ is independent of $(x,t)$. Since (see (\ref{mu1def}) and (\ref{mu2def}) for $t=0$)
\[
\Psi(x,0,k)=e^{-ik^2x\sig_3},\qquad \mbox{for }x\geq 0,
\]
\[
\Phi(x,0,k)=e^{-iBx\sig_3}E(k)e^{-ixX(k)\sig_3},\qquad \mbox{for }x\le 0,
\]
we conclude that
\be \label{Skdef}
S(k)=\Psi^{-1}(0,0,k)\Phi(0,0,k)=\Phi(0,0,k)=E(k).
\ee
Thus, we have
\be \label{Skdefforab}
S(k)=\left(\ba{cc}\bar a(\bar k)&b(k)\\-\bar b(\bar k)&a(k)\ea \right)=\left(\ba{cc}a(k)&b(k)\\b(k)&a(k)\ea \right),
\ee
where
\be \label{abdef}
\begin{split}
&a(k)=\bar a(\bar k)=\frac{1}{2}[\vphi(k)+\frac{1}{\vphi(k)}],\\
&b(k)=-\bar b(\bar k)=\frac{1}{2}[\vphi(k)-\frac{1}{\vphi(k)}].
\end{split}
\ee

\section{The basic Riemann-Hilbert problem}
The scattering relation (\ref{psiphirel}) involving the eigenfunctions $\Psi(x,t,k)$ and $\Phi(x,t,k)$ can be rewritten in the form of conjugation of boundary values of a piecewise analytic matrix-value function on a contour in the complex $k-$plane,namely:
\be \label{MMRHP}
M_+(x,t,k)=M_-(x,t,k)J(x,t,k),\qquad k\in \Sig,
\ee
where $M_\pm(x,t,k)$ denote the boundary vales of $M(x,t,k)$ according to a chosen orientation of $\Sig$, and $\Sig=\R \cup i\R \cup \gam \cup \bar \gam$.
\par
Indeed,let us write (\ref{psiphirel}) in the vector form:
\be \label{rewritepsiphirel}
\begin{split}
&\frac{\Phi^{(1)}(x,t,k)}{a(k)}=\Psi^{(1)}(x,t,k)+r(k)\Psi^{(2)}(x,t,k),\\
&\frac{\Phi^{(2)}(x,t,k)}{a(k)}=r(k)\Psi^{(1)}(x,t,k)+\Psi^{(2)}(x,t,k),
\end{split}
\ee
where
\be \label{rk}
r(k):=\frac{b(k)}{a(k)}=\frac{i}{kA}[k^2-B-\frac{A^2}{2}-X(k)],
\ee
and define the matrix $M(x,t,k)$ as follows:
\be \label{Mdef}
M(x,t,k)=\left\{\ba{cc}(\ba{cc}\frac{\Phi^{(1)}(x,t,k)}{a(k)}e^{it\tha(k)}&\Psi^{(2)}(x,t,k)e^{-it\tha(k)}\ea),&k\in \{k\in \C|\im k^2>0\backslash \gam\},\\(\ba{cc}\Psi^{(1)}(x,t,k)e^{it\tha(k)}&\frac{\Phi^{(2)}(x,t,k)}{a(k)}e^{-it\tha(k)}\ea),&k\in \{k\in \C|\im k^2<0\backslash \bar \gam\},\ea \right.
\ee
where
\be \label{thakdef}
\tha(k):=2k^4+\frac{x}{t}k^2,
\ee
Then the boundary values $M_+(x,t,k)$ and $M_-(x,t,k)$ relative to $\Sig$ are related by (\ref{MMRHP}),where
\be \label{jumpdef}
J(x,t,k)=\left\{\ba{lc} \left(\ba{cc}1-r^2(k)&-r(k)e^{-2it\tha(k)}\\r(k)e^{2it\tha(k)}&1\ea \right),&k^2\in \R,\\
                        \left(\ba{cc}1&0\\f(k)e^{2it\tha(k)}&1\ea \right),&k^2\in \gam,\\
                        \left(\ba{cc}1&f(k)e^{-2it\tha(k)}\\0&1\ea \right),&k^2\in \bar \gam,
                \ea
         \right.
\ee
with
\be \label{fkdef}
f(k):=r_+(k)-r_-(k).
\ee
\par
The jump relation (\ref{MMRHP}) considered together with the properties of the eigenfunctions listed in Proposition 1 suggests a way of representing the solution to the Cauchy problem (\ref{eq:DNLS}) and (\ref{eq:initial}) in terms of the solution of the Riemann-Hilbert problem, which is specified by the initial conditions (\ref{eq:initial}) via the associated spectral function $r(k)$.
\par
The solution $q(x,t)$ of the initial value problem (\ref{eq:DNLS}) and (\ref{eq:initial}) can be expressed in terms of the solution of the basic Riemann-Hilbert problem as follows:
\be \label{eqsol}
q(x,t)=2i\lim_{k \rightarrow \infty}(kM(x,t,k))_{12}.
\ee
where $M$ is the solution of the following Riemann-Hilbert problem:
\\
{\bf Basic Riemann-Hilbert problem \Rmnum{1}.}\\
Given $r(k),k^2\in \R$ and $f(k)=r_+(k)-r_-(k),k^2\in \gam \cup \bar \gam$, and $\Sig=\R \cup i\R \cup \gam \cup \bar \gam$, find a $2\times 2$ matrix-value function $M(x,t,k)$ such that
\begin{enumerate}
\item $M(x,t,k)$ is analytic in $k\in \C\backslash \Sig$.
\item $M(x,t,k)$ is bounded at the end points $E$ and $\bar E$.
\item The boundary value $M_\pm(x,t,k)$ at $\Sig$ satisfy the jump condition
       \[
       M_+(x,t,k)=M_-(x,t,k)J(x,t,k),\quad k\in \Sig 
       \]
      where the jump matrix $J(x,t,k)$ is defined in terms of $r(k)$ and $f(k)$ by (\ref{jumpdef}).
\item Behavior at $\infty$
      \[
      M(x,t,k)=\id+O(\frac{1}{k}),\qquad \mbox{as }k\rightarrow \infty.
      \]
\end{enumerate}
If we try to analysis the long-time asymptotic behavior of the GI-type of DNLS equation (\ref{eq:DNLS}) and (\ref{eq:initial}) with step-like initial value problem, this type of Riemann-Hilbert problem has a contradiction in the plane wave region. So we try to derive a new Riemann-Hilbert problem, which is similar to the type of nonlinear Schr\"odinger equation, to overcome this contradiction. That means we arrive at the following Riemann-Hilbert problem.
\par
We define
\be \label{Ndef}
N(x,t,k)=k^{-\frac{\hat \sig_3}{2}}M(x,t,k),
\ee
then the jump condition for $N$ is
\be \label{Njump}
N_+(x,t,k)=N_-(x,t,k)e^{-i(k^2x+2k^4t)\hat \sig_3}J_N(x,t,k).
\ee
introducing $\lam=k^2$ and control the branch of $k$ as $Sign\im k=Sign\im \lam$, and define the modified scattering data $\rho(\lam)=\frac{r(k)}{k}$, \cite{kn}.

\par
Then
\be \label{Xlam}
X(\lam)=\sqrt{(\lam-B-\frac{A^2}{2})^2+\lam A^2}=\sqrt{(\lam-B)^2+\frac{A^4}{4}+A^2B},
\ee
\be \label{Omlam}
\Om(\lam)=2(\lam+B)X(\lam).
\ee
and the segment
\be \label{Xvphiseglam}
\gam \cup \bar \gam:=\{\lam\in \C |\lam_1=B,\lam_2^2\le D^2 \},
\ee
where $\gam=\{k\in \C |\lam_1=B,\lam_2^2\le D^2,\im \lam_2>0\}$, $D^2=A^2B+\frac{A^4}{4}$, $\lam_1=\re{\lam}$ and $\lam_2=\im{\lam}$. Let $E=B+iD$, then $\gam=[E,B]$ and $\bar \gam=[B,\bar E]$.
And the jump condition for $N$ is
\be \label{NJdef}
N_+(x,t,\lam)=
N_-(x,t,\lam)e^{-i(\lam x+2\lam^2t)\hat \sig_3}J_N(x,t,\lam).
\ee
where
\be \label{jumpdefJN}
J_N(x,t,\lam)=\left\{\ba{ll}\left(\ba{cc}1-\lam \rho(\lam)^2&-\rho(\lam)e^{-2it\tha(\lam)}\\ \lam \rho(\lam)e^{2it\tha(\lam)}&1\ea \right),&\lam\in\R,\\
                            \left(\ba{cc}1&0\\\lam f(\lam)e^{2it\tha(\lam)}&1\ea \right),&\lam\in \gam,\\
                            \left(\ba{cc}1&f(\lam)e^{-2it\tha(\lam)}\\0&1\ea \right),&\lam\in \bar \gam,
\ea
\right.
\ee
where
\be
f(\lam)=\rho(\lam)_+-\rho(\lam)_-.
\ee
\begin{figure}[th]
\centering
\includegraphics{GI.2}
\caption{The oriented contour $\Sig=\R\cup\gam\cup\bar\gam$.}
\end{figure}
In other word,we have the following basic Riemann-Hilbert problem
\\
{\bf Basic Riemann-Hilbert problem \Rmnum{2}.}\\
Given $\rho(\lam),\lam\in \R$ and $f(\lam)=\rho(\lam)_+-\rho(\lam)_-,\lam\in \gam \cup \bar \gam$, and $\Sig=\R  \cup \gam \cup \bar \gam$, find a $2\times 2$ matrix-value function $N(x,t,\lam)$ such that
\begin{enumerate}
\item $N(x,t,\lam)$ is analytic in $\lam\in \C\backslash \Sig$.
\item $N(x,t,\lam)$ is bounded at the end points $E$ and $\bar E$.
\item The boundary value $N_\pm(x,t,\lam)$ at $\Sig$ satisfy the jump condition
       \[
       N_+(x,t,\lam)=N_-(x,t,\lam)J_N(x,t,\lam),\quad \lam\in \Sig \backslash \{ E,\bar{ E},B\},
       \]
      where the jump matrix $J_N(x,t,k)$ is defined in terms of $\rho(\lam)$ and $f(\lam)$ by (\ref{jumpdefJN}).
\item Behavior at $\infty$
      \[
      N(x,t,\lam)=\id+O(\frac{1}{\lam}),\qquad \mbox{as }\lam\rightarrow \infty.
      \]
\end{enumerate}

\section{Long-time Asymptotics}

The representation of the solution $q(x,t)$ of the initial value problem (\ref{DNLSandInit}) in terms of the solution of an associated basic Riemann-Hilbert problem allows using the ideas of the asymptotic analysis of oscillating Riemann-Hilbert problems \cite{dz,bv,diz,diz2,abmks1} for studying the long-time asymptotics of $q(x,t)$. The key fact leading to different asymptotics in different regions of the $(x,t)$ half-plane is that the behavior of the jump matrix of the basic Riemann-Hilbert problem as a function of the large parameter $t$ is different in these regions. Indeed, as seen on (\ref{jumpdefJN}), this behavior is governed by the sign of $\im \tha(\lam)$, which itself depends on $\x=\frac{x}{4t}$. As we have already written, three regions are to be distinguished:
\begin{enumerate}
\item A Zakharov-Manakov region:$\x>-B$.
\item A plane wave region:$\x<-\sqrt{2}D-B$.
\item An elliptic wave region:$-\sqrt{2}D-B<\x<-B$.
\end{enumerate}

\begin{figure}[th]
\centering
\includegraphics{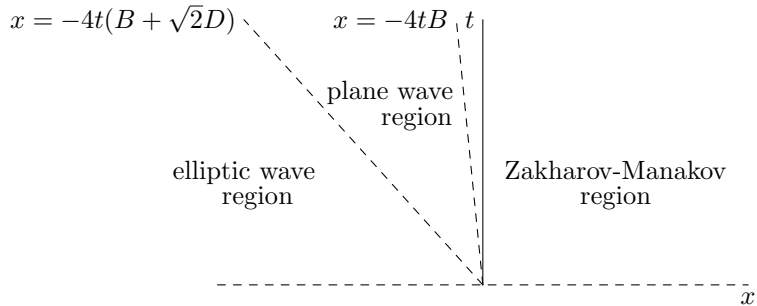}
\caption{The different regions of the $(x,t)-$plane.}
\end{figure}

\subsection{The Zakharov-Manakov region:$\x>-B$}

In this region $\x>-B$, we have $\im \tha(\lam)>0$ for all $\lam\in \gam$ and $\im \tha(\lam)<0$ for all $\lam\in \bar \gam$. Therefore, the exponentials in the jump matrix $J_N$, see (\ref{jumpdefJN}), are decaying as $t\rightarrow +\infty$ for $\lam\in \Sig\backslash \R$.
\par
This implies that one can follow the technique of asymptotic analysis proposed for the first time in \cite{dz}. The basic step of the procedure is a deformation of the original Riemann-Hilbert problem, with the help of the solution of an appropriate scalar Riemann-Hilbert problem, in order to obtain an equivalent Riemann-Hilbert problem whose jump matrix decays, in $t$, to a constant (in $\lam$) matrix. This leads to model Riemann-Hilbert problems whose solutions can be given explicitly.
\par
A particular feature of the Riemann-Hilbert problem under consideration is that the contour of the modified Riemann-Hilbert problem contains neither the real axis, where the jump matrix for the original Riemann-Hilbert problem oscillates with $t$, see (\ref{jumpdefJN}), nor the finite parts $\gam$ and $\bar \gam$. This happens due to the pure step-like initial conditions, which in turn implies that the associated spectral functions $\rho(\lam)$ and $\lam\rho(\lam)$ can be analytically extended from the contour to the whole $\lam$-plane.

\subsubsection{First transformation} \label{ZMfirst}

The first transform is as usual:
\be \label{ZSfirsttrans}
N^{(1)}(x,t,\lam)=N(x,t,\lam)\dta^{-\sig_3}(\lam),
\ee
where (\cite{xf})
\be \label{dta}
\dta(\lam)=\exp{\frac{1}{2\pi i}}\int_{-\infty}^{\lam_0}\frac{\log{(1-\lam' \rho(\lam')^2)}}{\lam'-\lam}d\lam',
\ee
is the solution of the following scalar Riemann-Hilbert problem:
\begin{itemize}
\item $\dta(\lam)$ is analytic in $\C \backslash (-\infty,\lam_0]$,
\item $\dta(\lam) \rightarrow 1$ as $\lam\rightarrow \infty$,
\item $\dta(\lam)$ satisfies the jump relation
      \be \label{dtaRHP}
      \dta_+(\lam)=\dta_-(\lam)(1-\lam\rho^2(\lam)),\qquad \lam\in (-\infty,\lam_0).
      \ee
\end{itemize}
Here, $\lam_0$ is the stationary point of the phase function $\tha(\lam)=2\lam^2+4\x\lam$, that is, $\tha'(\lam_0)=0$:
\[
\lam_0=-\x=\frac{-x}{4t}.
\]
Then $N^{(1)}(x,t,\lam)$ satisfies the jump condition
\be \label{ZSM1jump}
\begin{split}
&N_+^{(1)}(x,t,\lam)=N_-^{(1)}(x,t,N)J_N^{(1)}(x,t,\lam),\\
&\lam\in \Sig^{(1)}=\Sig,
\end{split}
\ee
where
\[
J_N^{(1)}(x,t,\lam)=\dta_-^{\sig_3}J_N\dta_+^{-\sig_3},
\]
that is
\be \label{ZMJ1}
J_N^{(1)}(x,t,\lam)=\left\{\ba{lc} e^{-it\tha \hat \sig_3}\left(\ba{cc}\frac{\dta_-}{\dta_+}(1-\lam\rho(\lam)^2)&-\rho\dta_+\dta_-\\\frac{\lam\rho}{\dta_+ \dta_-}&\frac{\dta_+}{\dta_-}\ea \right),&\qquad \lam\in \R,\\
                              \left(\ba{cc}\frac{\dta_-}{\dta_+}&0\\ \frac{\lam f}{\dta_+\dta_-}e^{2it\tha \sig_3}&\frac{\dta_+}{\dta_-}\ea \right),&\qquad \lam\in \gam,\\
                              \left(\ba{cc}\frac{\dta_-}{\dta_+}&f\dta_+\dta_-e^{-2it\tha \sig_3}\\0&\frac{\dta_-}{\dta_+}\ea \right),&\qquad \lam\in \bar \gam.\ea
\right.
\ee
From the Riemann-Hilbert problem of the $\dta$, we can find
\be \label{ZMj1def}
J_N^{(1)}(x,t,\lam)=\left\{ \ba{lc} e^{-it\tha \hat \sig_3}\left(\ba{cc}1-\lam \rho^2&-\rho\dta^2\\ \frac{\lam\rho}{\dta^2}&1\ea \right),&\qquad \lam>\lam_0,\\
e^{-it\tha \hat \sig_3}\left(\ba{cc}1&\frac{-\rho}{1-\lam\rho^2}\dta_-^2\\ \frac{\lam\rho}{1-\lam\rho^2} \frac{1}{\dta_+^2}&1-\lam\rho^2\ea \right),&\qquad \lam<\lam_0,\\
\left(\ba{cc} 1&0\\ \frac{\lam f}{\dta^2}e^{2it\tha}&1\ea \right),&\qquad \lam\in \gam,\\
\left(\ba{cc} 1&f\dta^2 e^{-2it\tha}\\0&1\ea \right),&\qquad \lam\in \bar \gam.\ea \right.
\ee

\subsubsection{Second transformation}\label{ZMscond}
The next transformation is:
\be \label{ZSsecondtrans}
N^{(2)}(x,t,\lam)=N^{(1)}(x,t,\lam)G(\lam),
\ee
where
\be \label{Gk}
G(\lam)=\left\{\ba{ll} \left(\ba{cc}1&\frac{\rho}{1-\lam\rho^2}\dta_-^2e^{-2it\tha}\\0&1\ea \right),&\qquad \lam\in D_1,\\
                    \left(\ba{cc}1&0\\\frac{\lam\rho}{1-\lam\rho^2} \frac{1}{\dta_+^2}e^{2it\tha}&1\ea \right),&\qquad \lam\in D_2,\\
                    \left(\ba{cc}1&-\rho\dta^2e^{-2it\tha}\\0&1\ea \right),&\qquad \lam\in D_3,\\
                    \left(\ba{cc}1&0\\\frac{-\lam\rho}{\dta^2}e^{2it\tha}&1\ea \right),&\qquad \lam\in D_4,\\
                    \id,&\qquad \lam\in D_5\cup D_6.
                    \ea
                    \right.
\ee
The domains $D_1,\ldots D_{6}$ are shown on the following Figure.
\begin{figure}[th]
\centering
\includegraphics{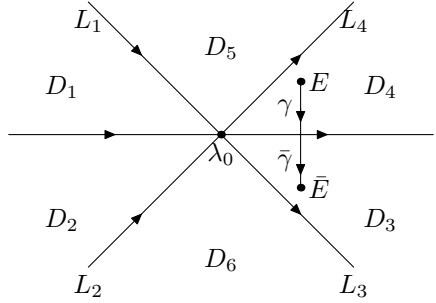}
\caption{The oriented contour $\Sig^{(2)}=L_1\cup L_2\cup L_3\cup L_4$.}
\end{figure}
\par
This new function $N^{(2)}$ solves the equivalent Riemann-Hilbert problem:
\[
\begin{split}
&N_+^{(2)}(x,t,\lam)=N_-^{(2)}(x,t,\lam)J_N^{(2)}(x,t,\lam),\\
&\lam\in \Sig^{(2)},
\end{split}
\]
where
\begin{sequation} \label{ZMJ2def}
J_N^{(2)}(x,t,\lam)=\left\{
\ba{ll}
\left(\ba{cc}1&\frac{-\rho}{1-\lam\rho^2}\dta_-^2e^{-2it\tha}\\0&1\ea \right),&\qquad \lam\in L_1,\\
\left(\ba{cc}1&0\\\frac{\lam\rho}{1-\lam\rho^2} \frac{1}{\dta_+^2}e^{2it\tha}&1\ea \right),&\qquad \lam\in L_2,\\
\left(\ba{cc}1&-\rho\dta^2e^{-2it\tha}\\0&1\ea \right),&\qquad \lam\in L_3,\\
\left(\ba{cc}1&0\\\frac{\lam\rho}{\dta^2}e^{2it\tha}&1\ea \right),&\qquad \lam\in L_4.
                      \ea
               \right.
\end{sequation}

\subsubsection{The last transformation}

Now $J_N^{(2)}(x,t,\lam)$ decays exponentially fast to the identity matrix, as $t\rightarrow +\infty$, and uniformly outside any neighborhood of $\lam=\lam_0$. Thus, we are in a situation where the asymptotic analysis of \cite{xf} works. Particularly,
\[
N^{(2)}(x,t,\lam)=Z(x,t,\lam)N^{as}(x,t,\lam),
\]
where $N^{as}(x,t,\lam)$ is a solution of the model problem explicitly given in terms of parabolic cylinder functions whereas $Z(x,t,\lam)$ can be estimated:
\[
Z(x,t,\lam)=\id +O(\frac{log t}{t^{\frac{1}{2}}}).
\]
Therefore, the final asymptotic result is as in \cite{xf} giving the main term of the asymptotic in terms of the modified reflection coefficient $\rho(\lam)$:

\begin{theorem}(The Zakharov-Manakov region)
In the region $x>-4tB$, the asymptotics, as $t\rightarrow +\infty$, of the solution $q(x,t)$ of the initial value problem (\ref{DNLSandInit}) is described by the Zakharov-Manakov type formula
\be
q(x,t)=q_{as}(x,t)+O(\frac{\log t}{t})
\ee
where
\be
\ba{l}
q_{as}=\frac{1}{\sqrt{t}}\alpha(\lam_0)e^{\frac{ix^2}{4t}-i\nu(\lam_0)\log t},\\
|\alpha(\lam_0)|^2=\frac{\nu(\lam_0)}{2}=-\frac{1}{4\pi}\log(1-\lam_0|\rho(\lam_0)|^2),\\
\arg \alpha(\lam_0)=-3\nu\log 2-\frac{\pi}{4}+\arg \Gam(i\nu)-\arg r(\lam_0)+\frac{1}{\pi}\int_{-\infty}^{\lam_0}\log|\lam-\lam_0|d\log(1-\lam|\rho(\lam)|^2),\\
\lam_0=-\frac{x}{4t}.
\ea
\ee
\end{theorem}

\subsection{The plane wave region: $\x<-\sqrt{2}D-B$}

For $x<-4t(B+\sqrt{2A^2(B+\frac{A^2}{4})})$, that means, $\im \tha(\lam)$ is negative on $\gam$ and positive on $\bar \gam$, which implies that the exponentials in (\ref{jumpdefJN}) increase with $t$. Thus, the jump matrix $J_N$ for the Riemann-Hilbert problem does not converge to a reasonable limit as $t\rightarrow \infty$.
\par
To bypass this difficulty, one deforms the Riemann-Hilbert problem in such a way that the phase $\im \tha(\lam)$ is replaced by another function, $g(\lam)$, providing suitable behavior of the modified jump matrix. The extension of the nonlinear steepest descent method for
Riemann-Hilbert problems, involving the "$g$-function mechanism" was first proposed by Deift, Venakides, and Zhou, see \cite{dvz1,dvz2}.
\par

\subsubsection{The $g$ function}

A natural choice for a $g$-function appropriate for the region adjacent to the half-axis $x < 0$, $t = 0$, is the phase appearing in the explicit expression for the eigenfunction $\Psi^p$, see (\ref{Psip}), associated with the "potential" $q^p$. Setting
\be \label{gdef}
g(x,t,\lam)=xX(\lam)+t\Om(\lam),
\ee
where $X(\lam)$ and $\Om(\lam)$ are defined in (\ref{Xlam}) and (\ref{Omlam}),we have
\be \label{Phipagain}
\Psi^p(x,t,k)=e^{i(\om t-Bx)\sig_3}E(\lam)e^{-ig(x,t,\lam)\sig_3}
\ee
The "signature table" for $\im g(\lam; \x)$ is the partition of the $\lam$-plane into maximal domains where the sign of $\im g(\lam; \x)$ is constant. Its form can be controlled by the zeros of the differential $dg(\lam)$. Indeed,
\be \label{dgk}
dg(\lam)=4\frac{(\lam-\mu_+)(\lam-\mu_-)}{X(\lam)}d\lam,
\ee
where
\be \label{mu+-}
\mu_\pm=\frac{B-\x}{2}\pm \sqrt{\frac{(B+\x)^2}{4}-\frac{\frac{A^4}{4}+A^2B}{2}},
\ee
Thus, for $\x<-(B+\sqrt{2A^2(B+\frac{A^2}{4})})$, $\mu_\pm$ are both real. Moreover,
\[
B<\mu_-<\mu_+<-\x.
\]
\par
In what follows the signature table of the function $\im g(\lam)$ for different values of $\x$ plays a very important role. The lines of separation between the different domains are the real  axile
\[
\lam_2=0,
\]
and the algebraic curve
\be \label{imgkcurve}
\lam_2^2(\lam_1+\x)=(\lam_1+B+2\x)[(\lam_1-B)(\lam_1+\x)+\frac{\frac{A^4}{4}+A^2B}{2}],
\ee
They are indeed given by $\im g(\lam)=0$. Because of
\[
\im g(\lam)=4\lam_2\{(\lam_1+B+2\x)[(\lam_1-B)(\lam_1+\x)+\frac{\frac{A^4}{4}+A^2B}{2}]-\lam_2^2(\lam_1+\x)\}
\]
\par
The equation (\ref{imgkcurve}) can be written:
\[
\lam_2^2(\lam_1+\x)=(\lam_1+B+2\x)[(\lam_1-\mu_+)(\lam_1-\mu_-)].
\]
And the signature table of the function $\im g(\lam)$ is shown in the following Figure 4.
\begin{figure}[th]
\centering
\includegraphics{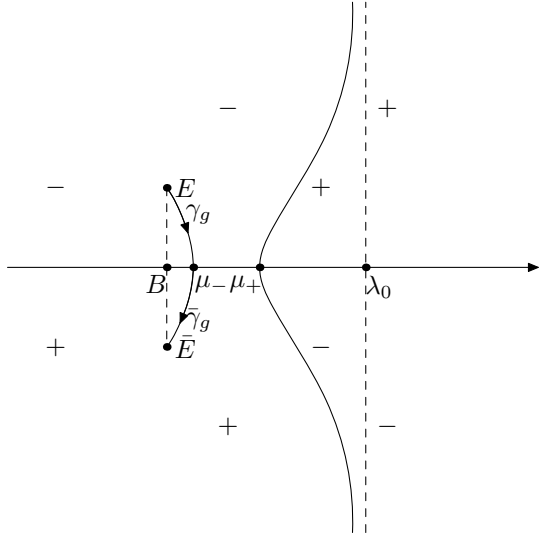}
\caption{The curves of $\im g(\lam)=0$ for $x<-4t(B+\sqrt{2A^2(B+\frac{A^2}{4})})$.}
\end{figure}
\par
The advantage of the signature table shown in Figure 4 is that there is a finite arc connecting the branch points $E$ and $\bar E$ such that $\im g(\lam)=0$ for all $\lam$ along this arc. Since the jump matrix depends on $t$ via exponentials of type $e^{\pm ig(\lam)}$, it is oscillatory along an arc where $\im g(\lam)=0$.
\par
This suggests to deform the original contour $\gam \cup \bar \gam$ of the basic Riemann-Hilbert problem to a new contour $\gam_g \cup \bar \gam_g$ which depends on $\x$ and where $\im g(\lam)=0$, and to view $X(\lam)$, thus also $g(\lam)$ as functions with branch cut $\gam_g \cup \bar \gam_g$.
\par
Another important feature of $g(\lam;\x)$ is that it has, up to a constant, the same large $\lam$ asymptotic behavior as the phase function $\tha(\lam)$:
\be \label{glargek}
g(\lam;\x)=t(2\lam^2+4\x \lam+g(\infty;\x))+O(\frac{1}{\lam}),\qquad \lam\rightarrow \infty,
\ee
where
\be \label{ginfty}
g(\infty;\x)=(\om-4B\x).
\ee

\subsubsection{The first transformation}

We put
\[
N^{(1)}(x,t,\lam)=e^{-itg(\infty,\x)\sig_3}N(x,t,\lam)e^{-i(\lam x+2\lam^2t-g(\lam))\sig_3},
\]
Then the matrix-value function $N^{(1)}(x,t,\lam)$ satisfies the following Riemann-Hilbert problem:
\[
N_+^{(1)}(x,t,\lam)=N_-^{(1)}(x,t,\lam)J_N^{(1)}(x,t,\lam),\qquad \lam\in \Sig^{(1)}=\R  \cup \gam_g \cup \bar \gam_g,
\]
with the jump matrix
\be \label{planeJ1def}
J_N^{(1)}(x,t,\lam)=\left\{\ba{lc} \left(\ba{cc}1-\lam\rho^2(\lam)&-\rho(\lam)e^{-2ig(\lam)}\\\lam\rho(\lam)e^{2ig(\lam)}&1\ea \right),&\qquad \lam\in \R,\\
                              \left(\ba{cc}e^{-2ig_-(\lam)}&0\\\lam f(\lam)&e^{2ig_-(\lam)}\ea \right),&\qquad \lam\in \gam_g,\\
                              \left(\ba{cc}e^{-2ig_-(\lam)}&f(\lam)\\0&e^{2ig_-(\lam)}\ea \right),&\qquad \lam\in \bar \gam_g.
                      \ea
               \right.
\ee
Here $g_\pm(\lam)$ are boundary values of $g$ on $\gam_g\cup \bar \gam_g$, and they are real. We also use the equation $g_+(\lam)=-g_-(\lam)$.

\subsubsection{The second transformation}

The next transformation is similar to the first transformation applied in the Zakharov–Manakov region, see Section \ref{ZMfirst}. It involves the solution $\dta(\lam)$ of the scalar Riemann-Hilbert problem \ref{dtaRHP}) but with $\mu_+$ instead of $\lam_0$,where $\mu_+$ is the stationary point of the new phase function $g(\lam)$. With this new scalar function $\dta(\lam)$, we set
\[
N^{(2)}(x,t,\lam)=N^{(1)}(x,t,\lam)\dta^{-\sig_3}(\lam),
\]
Then the matrix-value function $N^{(2)}(x,t,\lam)$ satisfies the following Riemann-Hilbert problem
\be \label{planeM2RHP}
N^{(2)}_+(x,t,\lam)=N^{(2)}_-(x,t,\lam)J_N^{(2)}(x,t,\lam),\qquad \lam\in \Sig^{(2)}=\Sig^{(1)},
\ee
where $J_N^{(2)}(x,t,\lam)$ is defined as follows:
\be \label{planeJ2def}
J_N^{(2)}(x,t,\lam)=\left\{ \ba{lc} e^{-ig \hat \sig_3}\left(\ba{cc}1-\lam\rho^2&-\rho\dta^2\\ \frac{\lam \rho}{\dta^2}&1\ea \right),&\qquad \lam>\mu_+,\\
e^{-ig \hat \sig_3}\left(\ba{cc}1&\frac{-\rho}{1-\lam \rho^2}\dta_-^2\\ \frac{\lam\rho}{1-\lam\rho^2} \frac{1}{\dta_+^2}&1-\lam\rho^2\ea \right),&\qquad \lam<\mu_+,\\
\left(\ba{cc} e^{-2ig_-(\lam)}&0\\ \frac{\lam f}{\dta^2}&e^{2ig_-(\lam)}\ea \right),&\qquad \lam\in \gam_g,\\
\left(\ba{cc} e^{-2ig_-(\lam)}&f\dta^2 \\0&e^{2ig_-(\lam)}\ea \right),&\qquad \lam\in \bar \gam_g.\ea \right.
\ee

\subsubsection{The third transformation}

The subsequent transformation
\[
N^{(3)}(x,t,\lam)=N^{(2)}(x,t,\lam)G(\lam),
\]
involves $G(\lam)$ defined similarly to (\ref{Gk}), with $t\tha$ replaced by $g$ and $\lam_0$ replaced by $\mu_+$. Then $N^{(3)}(x,t,\lam)$ satisfies the jump relation
\[
N_+^{(3)}(x,t,\lam)=N_-^{(3)}(x,t,\lam)J_N^{(3)}(x,t,\lam),
\]
across to the contour
\[
\Sig^{(3)}=L_1 \cup L_2\cup L_3\cup L_4 \cup \gam_g \cup \bar \gam_g,
\]
shown in Figure 5.
\begin{figure}[th]
\centering
\includegraphics{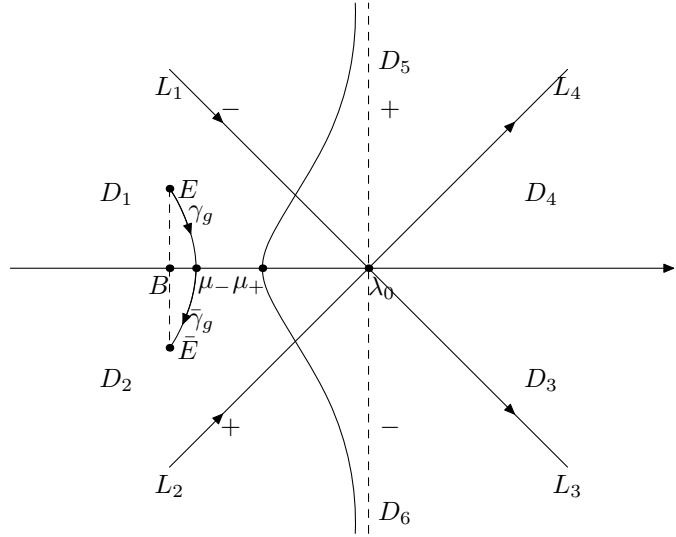}
\caption{The contour $\Sig^{(3)}=L_1\cup L_2\cup L_3\cup L_4\cup \gam_g\cup \bar\gam_g$ of the Riemann-Hilbert problem for $N^{(3)}$ for $x<-4t(B+\sqrt{2A^2(B+\frac{A^2}{4})})$.}
\end{figure}
\par
And we notice that
\par
1.For $\lam\in L_1\cup L_2\cup L_3\cup L_4$ the jump matrix $J_N^{(3)}(x,t,\lam)$ decays to the identity matrix, as $t\rightarrow \infty$, exponentially fast and uniformly outside any neighborhood of $\lam=\mu_+$.
\par
2.For $\lam\in  \gam_g$, the jump matrix $J_N^{(3)}(x,t,\lam)$ factorizes as
\begin{sequation}
\left(\ba{cc}1&(\frac{-\rho}{1-\lam\rho^2})_-\dta^{2}e^{-2ig_-(\lam)}\\0&1\ea \right)\left(\ba{cc}e^{-2ig_(\lam)}&0\\\lam f(\lam)\dta^{-2}(\lam)&e^{2ig_-(\lam)}\ea \right)\left(\ba{cc}1&(\frac{\rho}{1-\lam\rho^2})_+\dta^{2}e^{2ig_-(\lam)}\\0&1\ea \right)
\end{sequation}
\par
3.For $\lam\in \bar \gam_g$, the jump matrix $J_N^{(3)}(x,t,\lam)$ factorizes as
\begin{sequation}
\left(\ba{cc}1&0\\(\frac{-\lam\rho}{1-\lam\rho^2})_-\dta^{-2}e^{2ig_-(\lam)}&1\ea \right)\left(\ba{cc}e^{-2ig_-(\lam)}&f(\lam)\dta^2(\lam)\\0&e^{2ig_-(\lam)}\ea \right)\left(\ba{cc}1&0\\(\frac{\lam\rho}{1-\lam\rho^2})_+\dta^{-2}e^{2ig_-(\lam)}&1\ea \right)
\end{sequation}
\par
4.Using the identities
\[
1+\lam f (\frac{-\rho}{1-\lam\rho^2})_-=0,
\]
\[
1+ f (\frac{\lam\rho}{1-\lam\rho^2})_+=0,
\]
we find
\be \label{planeJ3def}
J_N^{(3)}(x,t,k)=\left\{\ba{lc} \left(\ba{cc}0&-(\lam f)^{-1}(\lam)\dta^2(\lam)\\\lam f(\lam)\dta^{-2}(\lam)&0\ea \right),&\qquad \lam\in \gam_g,\\
                              \left(\ba{cc}0&f(\lam)\dta^2(\lam)\\-f^{-1}(\lam)\dta^{-2}(\lam)&0\ea \right),&\qquad \lam\in \bar \gam_g,
                      \ea
               \right.
\ee
In order to arrive at a Riemann-Hilbert problem whose jump matrix does not depend on $\lam$, we introduce a factorization involving a scalar function $F(\lam)$ to be defined;
\be \label{planeJ3fac}
J_N^{(3)}(x,t,\lam)=\left(\ba{cc}F_+^{-1}(\lam)&0\\0&F_+(\lam)\ea \right)\left(\ba{cc}0&i\\i&0\ea \right)\left(\ba{cc}F_-(\lam)&0\\0&F_-^{-1}(\lam)\ea \right),
\ee
in such a way that the boundary values $F_\pm(\lam)$ of $F(\lam)$ along the two sides of $\gam_g\cup \bar \gam_g$ satisfy
\be \label{FkRHP}
F_-(\lam)F_+(\lam)=\left\{\ba{lc} -i\lam f(\lam)\dta^{-2}(\lam)&\qquad  \lam\in \gam_g,\\i f^{-1}(\lam)\dta^{-2}(\lam)&\qquad  \lam\in \bar \gam_g.\ea \right.
\ee
Indeed, once (\ref{planeJ3fac}) is satisfied, one can absorb the diagonal factors into a new piecewise analytic function whose jump across $\gam_g\cup \bar \gam_g$ is only the constant middle factor in (\ref{planeJ3fac}).
\par
Thus, we arrive at the following scalar Riemann-Hilbert problem:\\
{\bf Scalar Riemann-Hilbert problem.}\\
Find a scalar function $F(\lam)$ such that
\begin{itemize}
\item $F(\lam)$ and $F^{-1}(\lam)$ are analytic in $\C \backslash \{\gam_g\cup \bar \gam_g\}$.
\item $F(\lam)$ satisfies the jump relation:
\be \label{FkscalarRHP}
F_+(\lam)F_-(\lam)=\left\{\ba{lc} -i\lam f(\lam)\dta^{-2}(\lam)=a_+^{-1}(\lam)a_-^{-1}(\lam) \sqrt{\lam}\dta^{-2}(\lam),&\qquad \lam\in \gam_g,\\i f^{-1}(\lam)\dta^{-2}(\lam)=a_+(\lam)a_-(\lam) \sqrt{\lam}\dta^{-2}(\lam),&\qquad \lam\in \bar \gam_g.\ea \right.
\ee
where the contour $\gam_g\cup \bar \gam_g$ is oriented from $E$ to $\bar E$, and
\item $F(\lam)$ is bounded at $\lam=\infty$.
\end{itemize}
Introducing
\be \label{Hkdef}
H(\lam)=\left\{\ba{lc} F(\lam)a(\lam),&\qquad \lam\in \C_+\backslash \gam_g,\\
                       \frac{F(\lam)}{a(\lam)},&\qquad \lam\in \C_-\backslash \bar \gam_g.\ea \right.
\ee
then the jump relation (\ref{FkscalarRHP}) transforms to
\be \label{logHk}
[\frac{\log {H(\lam)}}{X(\lam)}]_+-[\frac{\log{H(\lam)}}{X(\lam)}]_-=
\left\{\ba{ll}\frac{\log {\sqrt{\lam}\dta^{-2}(\lam)}}{X(\lam)_+},&\qquad \lam\in \gam_g\cup \bar \gam_g,\\
              \frac{\log{a^2(\lam)}}{X(\lam)},&\qquad \lam\in \R.
       \ea
\right.
\ee
The Sokhotski-Plemelj formula shows that this last jump relation is satisfied by
\be \label{Hkpre}
H(k)=\exp\{\frac{X(\lam)}{2\pi i}[\int_{\gam_g\cup \bar \gam_g}\frac{\log{\sqrt{s}}+\log{\dta^{-2}(s,\x)}}{s-\lam}\frac{ds}{X_{+}(s)}+\int_{\R}\frac{\log ab(s)}{s-\lam}\frac{ds}{X(s)}]\}
\ee
Then $F(\lam)$ is defined in terms of $H(\lam)$ by (\ref{Hkdef}). At $\lam=\infty$ we find
\[
F(\infty)=H(\infty)=e^{i\phi(\x)},
\]
where
\be \label{phipre}
\phi(\x)=\frac{1}{2\pi}[\int_{\gam_g\cup \bar \gam_g}\frac{\log{\sqrt{s}\dta^{-2}(s,\x)}}{X_{+}(s)}ds+\int_{\R}\frac{\log a^2(s)}{X(s)}ds]
\ee
with
\be
\dta(\lam,\x)=\exp{\frac{1}{2\pi i}}\int_{-\infty}^{\mu_+}\frac{\log{(1-\lam' \rho(\lam')^2)}}{\lam'-\lam}d\lam',
\ee
Using the relation $1-\lam\rho^2(\lam)=a^{-2}(\lam)$, we find a simpler expression for $\phi(\x)$:
\[
\phi(\x)=\frac{1}{2\pi}[\int_{\mu_+}^{+ \infty}\log{a^2(\lam)}\frac{d\lam}{X(\lam)}+\int_{\gam_g\cup \bar \gam_g}\frac{\log{\sqrt{\lam}}}{X_{+}(\lam)}d\lam]
\]

\subsubsection{The fourth transformation}

The factorization (\ref{planeJ3fac}) suggests a fourth transformation
\[
N^{(4)}(x,t,\lam)=F^{\sig_3}(\infty,\x)N^{(3)}(x,t,\lam)F^{-\sig_3}(\lam,\x),
\]
Then we have
\[
N^{(4)}_+(x,t,\lam)=N^{(4)}_-(x,t,\lam)J_N^{(4)}(x,t,\lam)
\]
For $\lam\in \gam_g\cup \bar \gam_g$ the jump matrix $J_N^{(4)}(x,t,\lam)$ is constant
\[
J_N^{(4)}(x,t,\lam)=J_N^{mod}=\left(\ba{cc}0&i\\i&0\ea \right).
\]
1.For $\lam\in \gam_g\cup \bar \gam_g$ the jump matrix $J_N^{(4)}(x,t,\lam)$ is constant:
\[
J_N^{(4)}(x,t,\lam)=J_N^{mod}=\left(\ba{cc}0&i\\i&0\ea \right).
\]
2.For $\lam\in L\cup \bar L$, the jump matrix $J_N^{(4)}(x,t,\lam)$ decays to the identity
\[
J_N^{(4)}(x,t,\lam)=\id + O(\frac{1}{e^{\varplon t}}).
\]

\subsubsection{The final transformation}

Finally, we can express $N^{(4)}$ in the form
\[
N^{(4)}(x,t,\lam)=N^{err}(x,t,\lam)N^{mod}(x,t,\lam),
\]
where $N^{mod}(x,t,\lam)$ solves the model problem:
\be \label{planemodRHP}
N_-^{mod}(x,t,\lam)=N_+^{(mod)}(x,t,\lam)J_N^{mod},\qquad \lam\in \gam_g\cup \bar \gam_g,
\ee
with constant jump matrix
\[
J_N^{mod}=\left(\ba{cc}0&i\\i&0\ea \right),
\]
and $N^{err}(x,t,\lam)=\id+O(t^{-\frac{1}{2}})$.

\par
As for the model problem, since $\vphi(\lam)_-=i\vphi(\lam)_+$ on $\gam_g\cup \bar \gam_g$, its solution can be given explicitly in terms of $\vphi(\lam)$:
\[
N^{mod}(x,t,\lam)=\frac{1}{2}\left(\ba{cc}\vphi(\lam)+\frac{1}{\vphi(\lam)}&\vphi(\lam)-\frac{1}{\vphi(\lam)}\\\vphi(\lam)-\frac{1}{\vphi(\lam)}&\vphi(\lam)+\frac{1}{\vphi(\lam)}\ea \right).
\]

\subsubsection{Back to the original problem}

Let $N^{*}(x,t,\lam)$, $*=$ (1),(2),(3),(4),mod, denote the solution of the Riemann-Hilbert problem $RH^{*}$, and let
\[
m_{12}^{*}(x,t)=\lim_{\lam\rightarrow \infty}(\lam M^{*}(x,t,\lam))_{12},
\]
Then, going back to the determination of $q(x,t)$ in terms of the solution of the basic Riemann-Hilbert problem, we have
\be \label{planebackq}
\begin{split}
q(x,t)=&2im(x,t)_{12}=2ie^{2ig(\infty,\x)}m^{(1)}(x,t)_{12}\\
=&2ie^{2ig(\infty,\x)}m^{(2)}(x,t)_{12}+O(t^{-\frac{1}{2}})\\
=&2ie^{2ig(\infty,\x)}m^{(3)}(x,t)_{12}+O(t^{-\frac{1}{2}})\\
=&2ie^{2ig(\infty,\x)}m^{(4)}(x,t)_{12}F^{-2}(\infty,\x)+O(t^{-\frac{1}{2}})\\
=&2ie^{2ig(\infty,\x)}m^{mod}(x,t)_{12}F^{-2}(\infty,\x)+O(t^{-\frac{1}{2}}).
\end{split}
\ee
Taking into account that $g(\infty,\x)=\om t-4Bx$, $2im^{mod}(x,t)_{12}=A$ and $F^{-2}(\infty,\x)=e^{-2i\phi(\x)}$we arrive at the following theorem:

\begin{theorem}({\bf Plane wave region})
In the region $x<-4t(B+\sqrt{2A^2(B+\frac{A^2}{4})})$,the asymptotics, as $t\rightarrow +\infty$, of the solution $q(x,t)$ of the initial value problem (\ref{DNLSandInit}) takes the form of a plane wave:
\be \label{planefinalq}
q(x,t)=Ae^{2i(\om t-Bx-\phi(\x))}+O(t^{-\frac{1}{2}}),\qquad t\rightarrow +\infty.
\ee
\end{theorem}

\begin{remark}
If we let $\x\rightarrow +\infty$, then $\mu_+\rightarrow +\infty$, then $\phi(\x)\rightarrow \phi$, with $\phi=\frac{1}{2\pi}\int_{\gam_g\cup \bar \gam_g}\frac{\log{\sqrt{\lam}}}{X_{+}(\lam)}d\lam$, and then the above equation (\ref{planefinalq}) reduce to $q(x,t)=Ae^{2i(\om t-Bx-\phi)}$, this is correspondence to our initial condition up to a phase shift.
\end{remark}

\subsection{The elliptic region:$-4t(B+\sqrt{2}D)<x<-4tB$}

For the limit case $\x_0=-(B+\sqrt{2A^2(B+\frac{A^2}{4})})$, we have $\mu_+(\x_0)=\mu_-(\x_0)$, see Figure 7, whereas for $\x>-(B+\sqrt{2A^2(B+\frac{A^2}{4})})$, $\mu_+$ and $\mu_-$ become non-real, complex conjugated numbers. As a result, the $g$-function mechanism with $g(\lam;\x)$ as in the plane wave region fails. This shows that there is a break in the qualitative picture of the asymptotic behavior at $\x=\x_0$.

\subsubsection{The new $g$-function}
A suitable $g$-function for $\x>-(B+\sqrt{2A^2(B+\frac{A^2}{4})})$ can be obtained as follows. First, we need to introduce a new real stationary point $\mu(\x)$ which must be a zero of the new differential $d\hat g$. On the other hand we have to preserve the asymptotic behavior of the $g$-function for large $\lam$. To do so we must change the denominator of the differential $d\hat g$. Thus the new differential takes the form:
\be \label{newg-func}
d\hat g(\lam,\x)=4\frac{(\lam-\mu(\x))(\lam-\mu_-(\x))(\lam-\mu_+(\x))}{\sqrt{(\lam-E)(\lam-\bar E)(\lam-d(\x))(\lam-\bar d(\x))}}d\lam,
\ee
where $\mu(\x),\mu_\pm(\x)$, and $d(\x),\bar d(\x)$ are to be determined.
\par
If $\mu=d=\bar d$, then the new differential coincides with the previous one, that is $dg=d\hat g$, which is expected to hold for the value $\x_0$ of $\x$ limiting the two adjacent asymptotic regions.
\par
\par%
Now we consider $d\hat g$ as an Abelian differential of the second kind with poles at $\infty_\pm$ on the Riemann-Hilbert surface of
\[
\om(\lam)=\sqrt{(\lam-E)(\lam-\bar E)(\lam-d(\x))(\lam-\bar d(\x))},
\]
with
\[
E=B+iD,\quad d(\x)=d_1(\x)+id_2(\x)
\]
The branch of the square root is fixed by the asymptotics on the upper sheet:
\[
\om(\lam)=\lam^2+O(\lam),\qquad \lam\rightarrow \infty_+.
\]
We choose on this Riemann surface a basis $\{a,b\}$ of cycles as follows. The $b$-cycle is a closed clock-wise oriented simple loop around the arc $\gam_{E,d}$ joining $E$ and $d$. The $a$-cycle starts on the upper sheet from the left side of the cut $\gam_{E,d}$, goes to the left side of the
cut $\gam_{\bar d,\bar E}$, proceeds to the lower sheet, and then returns to the starting point.
\par
We can also write the Abelian differential $d\hat g(\lam)$ in the form:
\be \label{dhatg}
d\hat g(\lam)=4\frac{\lam^3+c_2\lam^2+c_1\lam+c_0}{\om(\lam)}d\lam,
\ee
and normalize it so that its $a-$period vanishes. This determines $c_0$:
\[
c_0=-\frac{\int_{\bar d}^{d}(\lam^3+c_2\lam^2+c_1\lam)\frac{d\lam}{\om(\lam)}}{\int_{\bar d}^{d}\frac{d\lam}{\om(\lam)}}\in \R.
\]
We also require that $\hat g(\lam)$ has the same large-$\lam$ behavior as the original phase function $\tha(\lam)$:
\[
\hat g(\lam)=2\lam^2t+4\lam x+O(1),\qquad \lam\rightarrow \infty_+.
\]
This condition implies
\[
c_1=(B-\x)d_1-B\x+\frac{1}{2}(d_2^2+D^2),
\]
\[
c_2=\x-B-d_1,
\]
Define $\hat g(\lam)$ as the sum of two Abelian integrals:
\be
\hat g(\lam,\x)=2(\int_{E}^{\lam}+\int_{\bar E}^{\lam})\frac{\lam^3+c_2\lam^2+c_1\lam+c_0}{\om(\lam)}d\lam.
\ee
Then it evidently has real $b-$period
\be \label{Bhatg}
B_{\hat g}=2(\int_{E}^{d}+\int_{\bar E}^{\bar d})\frac{\lam^3+c_2\lam^2+c_1\lam+c_0}{\om(\lam)}d\lam.
\ee
Now notice that $\hat g(\lam)$ can be written as a single Abelian integral
\[
\hat g(\lam)=4\int_{E}^{k}\frac{\lam^3+c_2\lam^2+c_1\lam+c_0}{\om(\lam)}d\lam
\]
and indeed
\[
B_{\hat g}=\int_{b}d\hat g.
\]
The large-$\lam$ asymptotics of $\hat g(\lam,\x)$ can now be specified as
\[
\hat g(\lam,\x)=2\lam^2t+4\x\lam t+\hat g(\infty,\x)+O(\lam^{-1}).
\]
where
\be
\hat g(\infty,\x)=t(2(\int_{E}^{\infty}+\int_{\bar E}^{\infty})[\frac{\lam^3+c_2\lam^2+c_1\lam+c_0}{\om(\lam)}-(\lam+\x)]d\lam+2D^2-2B^2-4B\x)
\ee
is a real function of $\x$.

\begin{remark}
For $\x=-B$, if we set $\mu(-B)=d_1(-B)=B$ and $d_2(-B)=D$, that is, $d(-B)=E$ and $\bar d(-B)=\bar E$, then $\hat g(\lam,-B)$ coincide(up to a constant) with $\tha(\lam,-B)$:
\[
\hat g(\lam,-B)=\tha(\lam,-B)+2|E|^2.
\]
which provides matching at the interface with the Zakharov-Manakov region.
\end{remark}
\par
In order to define $\mu,\mu_{\pm}$ and $d$ as functions of $\x$, let us compare the forms (\ref{newg-func}) and (\ref{dhatg}) of the differential $d\hat g$. This gives $(\mu_{\pm}=\mu_1\pm i\mu_2):$
\begin{equation*}
\ba{l}
\mu+2\mu_1-d_1=B-\x,\\
2\mu\mu_1+\mu_1^2+\mu_2^2+(\x-B)d_1-\frac{1}{2}d_2^2=\frac{1}{2}D^2-B\x,\\
\mu(\mu_1^2+\mu_2^2)=-c_0(\x,d_1,d_2).
\ea
\end{equation*}
The local expansion of $\hat g(\lam)$ at $\lam=d$ is of the form
\[
\hat g(\lam)=B_{\hat g}+g_1(\lam-d)^{1/2}+g_2(\lam-d)^{3/2}+\cdots,
\]
where $B_{\hat g}$ is real. The signature table for $\im \hat g(\lam)$ must have three branches of the curve $\im \hat g(\lam)=0$ going out from the point $d$, see Figure 6. Indeed:
\begin{itemize}
\item Since $\hat g(E)=0$, one branch should connect $d$ with $E$.
\item There should exist a branch separating the basins of $+$ and $-$ near the real axis.
\item Since $\hat g(\lam)$ behaves like $\tha(\lam)$ for large $\lam$, there should be an infinite branch going to infinity along the asymptotic line $\re \lam=-\x$.
\end{itemize}
\begin{figure}[th]
\centering
\includegraphics{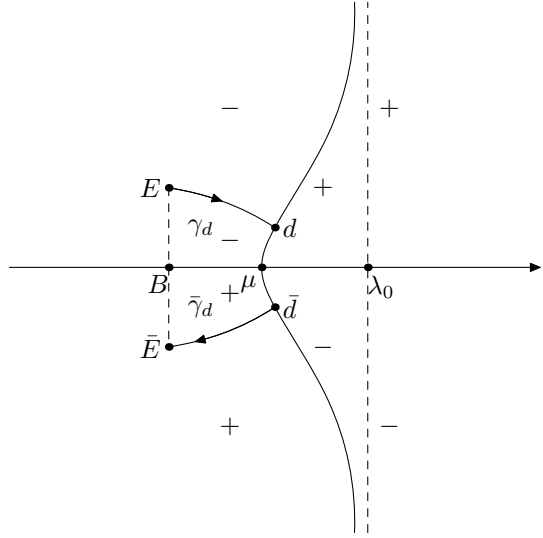}
\caption{The curves of $\im \hat g(\lam)=0$ for $-4t(B+\sqrt{2A^2(B+\frac{A^2}{4})})<x<-4tB$.}
\end{figure}
Therefore, we arrive at the requirement $g_1=0$, that is
\[
(\lam-d)^{1/2}\hat g'(\lam)|_{\lam=d}=4\frac{(d-\mu(\x))(d-\mu_-(\x))(d-\mu_+(\x))}{\sqrt{(\lam-E)(\lam-\bar E)(d-\bar d)}}=0
\]
The fact that $\mu$ is real implies that $\mu_+=d$ and $\mu_-=\bar d$, which finally leads to the following ansatz for $d\hat g(\lam)$:
\[
d\hat g(\lam)=4(\lam-\mu(\x))\sqrt{\frac{(\lam-d(\x))(\lam-\bar d(\x))}{(\lam-E)(\lam-\bar E)}}d\lam,
\]
where $\mu(\x),d_1(\x)$ and $d_2(\x)$ ($d=d_1+id_2,d_2\ge 0$) satisfy the equations:
\begin{subequations}\label{mud2intBsys}
\be \label{mu}
\mu=B-\x-d_1,
\ee
\be \label{d2}
d_2^2=D^2-2(B-\mu)(B-d_1),
\ee
\be \label{intB}
\int_{B-iD}^{B+iD}\sqrt{\frac{(\lam-d_1)^2+d_2^2}{(\lam-B)^2+D^2}}(\lam-\mu)d\lam=0.
\ee
\end{subequations}
Recall that (\ref{mu}) and (\ref{d2}) follow from the requirement that
\[
d\hat g(\lam)=(4\lam+4\x+O(\lam^{-2}))d\lam,\qquad \mbox{as }\lam\rightarrow \infty.
\]
while (\ref{intB}) is the normalization condition $\int_{\bar E}^{E}d\hat g(\lam)=0$.
\par
Substituting (\ref{mu}) and (\ref{d2}) into (\ref{intB}) yields an equation relating implicitly $d_1$ and $\x$. In terms of the variables $u$ and $v$, where
\[
u=\frac{B-d_1}{D},\qquad v=\frac{\x+B}{2D}.
\]
this equation reads
\be \label{Fuv}
\cF(u,v)=\int_{-1}^{1}\sqrt{\frac{(i\tau+1)^2+1-4uv+2u^2}{1-\tau^2}}(i\tau+2v-u)d\tau=0.
\ee
which is considered for $0\le v\le \frac{\sqrt{2}}{2}$ and $u\ge 0$. It is easy to check that $\cF(0,v)=4v$(and thus $\cF(0,v)>0$ for $v>0$), $\cF(+\infty,v)<0,\cF(0,0)=\cF(\frac{\sqrt{2}}{2},\frac{\sqrt{2}}{2})=0$ and $\cF_{u}(u,v)<0 $ for $(u,v)\ne (\frac{\sqrt{2}}{2},\frac{\sqrt{2}}{2})$. Therefore, (\ref{Fuv}) determines a unique function $u=u(v),v\in[0,\frac{\sqrt{2}}{2}]$ such that $u(0)=0$ and $u(\frac{\sqrt{2}}{2})=\frac{\sqrt{2}}{2}$. Consequently, we have that the system (\ref{mud2intBsys}) determines uniquely $d_1(\x),d_2(\x)$ and $\mu(\x)$, such that $d_1(-B-\sqrt{2}D)=B+\sqrt{2}D$ and $d_1(-B)=B$.
\par
We have now specified a $g-$function $\hat g(\lam)$ whose signature table is as in Figure 8. Hence, we can begin deforming the basic Riemann-Hilbert problem.

\subsubsection{The first deformation}

We deform the part $\gam \cup \bar \gam$ of the contour of the basic Riemann-Hilbert problem into a contour $\gam_{E,\bar E}$ connecting $E$ and $\bar E$ in such a way that it contains:
\begin{enumerate}
\item Two arcs $\gam_d$ and $\bar \gam_d$ connecting, respectively, $E$ with $d$ and $\bar d$ and $\bar E$, and where $\im \hat g(\lam)=0$;
\item An arc $\gam_{\mu}$ connecting $d$ and $\bar d$, passing through $\mu$, and along which $\im \hat g(\lam)<0$ for $\im \lam<0$ and $\im \hat g(\lam)>0$ for $\im \lam>0$.
\end{enumerate}
\begin{figure}[th]
\centering
\includegraphics{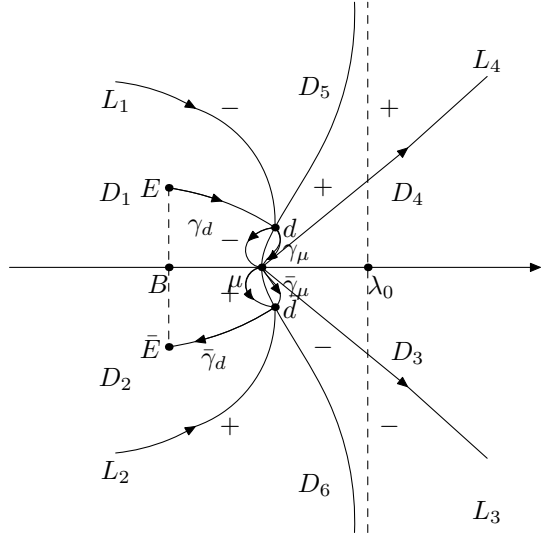}
\caption{The contour $\Sig^{(3)}=L_1\cup L_2\cup L_3\cup L_4\cup\gam_d\cup \bar\gam_d\cup\gam_{\mu}\cup\bar \gam_{\mu}$ for $-4t(B+\sqrt{2A^2(B+\frac{A^2}{4})})<x<-4tB$.}
\end{figure}
Supplying $\gam_{E,\bar E}=\gam_{\mu}\cup \gam_d \cup \bar \gam_{ d}$ with the orientation as going from $E$ to $\bar E$, we fix the branch of $\hat g(\lam)$ as having a jump across $\gam_{E,\bar E}$:
\[
\ba{ll}\hat g(\lam)_++\hat g(\lam)_-=0,&\qquad \lam\in \gam_d\cup \bar \gam_d;\\
       \hat g(\lam)_+-\hat g(\lam)_-=B_{\hat g},&\qquad \lam\in \gam_{\mu},\\
       \mbox{with $\im B_{\hat g}=0$}&
\ea
\]

\subsubsection{The second transformation}

The further series of transformations
\[
N(x,t,\lam)\leadsto N^{(1)}(x,t,\lam)\leadsto N^{(2)}(x,t,\lam)\leadsto N^{(3)}(x,t,\lam)
\]
is similar to that for the plane wave region but
\begin{enumerate}
\item with $g(\lam)$ replaced by $\hat g(\lam)$,
\item with $\mu$, which is the real stationary point of $\hat g(\lam)$ instead of $\mu_+$,
\item with the partition into domains with boundaries $L$ as shown in Figure 7.
\end{enumerate}
The jump matrix $J_N^{(3)}(x,t,\lam)$ is as follows:
\begin{itemize}
\item For $\lam\in L_j$ at a fixed positive distance from the stationary point $\lam=\mu(\x)$,
\[
J_N^{(3)}(x,t,\lam)=\id+O(e^{-\varplon t})\mbox{  as }t\rightarrow +\infty.
\]
\item For $\lam\in \gam_{\mu}$ we have
\be \label{ellJ3gammu}
J_N^{(3)}(x,t,\lam)=\left\{\ba{lc} \left(\ba{cc}e^{-itB_{\hat g}}&0\\\lam f(\lam)\dta^{-2}(\lam)e^{it(\hat g_+(\lam)+\hat g_-(\lam))}&e^{itB_{\hat g}}\ea \right),&\quad \im \lam>0,\\
\left(\ba{cc}e^{-itB_{\hat g}}&f(\lam)\dta^{2}(\lam)e^{-it(\hat g_+(\lam)+\hat g_-(\lam))}\\0&e^{itB_{\hat g}}\ea \right),&\quad \im \lam<0,
\ea \right.
\ee
Thus, away from $d$,$\mu$ and $\bar d$ and as $t\rightarrow +\infty$, $J_N^{(3)}(x,t,\lam)$ is close to a diagonal matrix:
\be \label{ellJ3gammuasy}
J_N^{(3)}(x,t,\lam)=\left(\ba{cc}e^{-itB_{\hat g}}&0\\0&e^{itB_{\hat g}}\ea \right)+O(e^{-\varplon t}),\qquad t\rightarrow +\infty.
\ee
\item For $\lam\in \gam_d\cup \bar \gam_d$, similarly to the plane wave region, $J_N^{(3)}(x,t,\lam)$ reduces to
\be \label{ellJ3gamd}
J_N^{(3)}(x,t,\lam)=\left\{\ba{lc} \left(\ba{cc}0&-f^{-1}(\lam)\dta^2(\lam)\\\lam f(\lam)\dta^{-2}(\lam)&0\ea \right),&\qquad \lam\in \gam_d,\\
                              \left(\ba{cc}0&f(\lam)\dta^2(\lam)\\-\lam f^{-1}(\lam)\dta^{-2}(\lam)&0\ea \right),&\qquad \lam\in \bar \gam_d,
                      \ea
               \right.
\ee
\end{itemize}
In order to arrive at a Riemann-Hilbert problem with a jump matrix independent of $\lam$, we proceed as in the plane wave region.
\par
{\bf Scalar Riemann-Hilbert problem.}
We are looking for a scalar function $F(\lam)$ analytic in $\C\backslash \gam_d\cup \bar \gam_d$ such that
\be \label{ellFkscalarRHP}
F_-(\lam)F_+(\lam)=h(\lam)\sqrt{\lam}\dta^{-2}(\lam),\qquad \lam\in \gam_d\cup \bar \gam_d,
\ee
where
\be \label{hdef}
h(\lam)=\left\{\ba{lc} -i\sqrt{\lam}f(\lam),&\qquad \lam\in \gam_g,\\i \sqrt{\lam}^{-1}f^{-1}(k),&\qquad \lam\in \bar \gam_g.\ea \right.
\ee
\par
After solving this scalar problem, $J_N^{(3)}(x,t,\lam)$ can be factorized as in (\ref{planeJ3fac}). This factorization allows absorbing the diagonal factors into a new Riemann-Hilbert problem with constant jump matrix on $\gam_d\cup \bar \gam_d$.
\par
However, an important difference with the plane wave region is that now the jump conditions (\ref{ellFkscalarRHP}) for $F(\lam)$ are specified on two disjoint arcs. This implies that in order to arrive at a jump condition in additive form, we are led to use
\[
\om(\lam)=\sqrt{(\lam-E)(\lam-\bar E)(\lam-d(\x))(\lam-\bar d(\x))}
\]
Indeed, (\ref{ellFkscalarRHP}) can be rewritten as
\be \label{elllogFRHP}
[\frac{\log{F(\lam)}}{\om(\lam)}]_+-[\frac{\log{F(\lam)}}{\om(\lam)}]_-=\frac{\log{h(\lam)}}{\om_{+}(\lam)},\quad \lam\in \gam_d \cup \bar \gam_d,
\ee
and thus for $F(\lam)$, we have
\be \label{ellFk}
F(\lam)=\exp\{\frac{\om(\lam)}{2\pi i}\int_{\gam_d\cup \bar \gam_d}\frac{\log{h(s)}}{\om_{+}(s)}\frac{ds}{s-\lam}\}
\ee
But now $F(\lam)$ has an essential singularity at infinity:
\[
F(\lam)=F_{\infty}e^{i\Dta \lam}(1+O(\lam^{-1})),\qquad \lam\rightarrow \infty.
\]
where
\be\label{Dta}
\Dta=\Dta(\x)=\frac{1}{2\pi}\int_{\gam_d\cup \bar \gam_d}\frac{\log{h(\lam)}}{\om_{+}(\lam)}d\lam.
\ee
and
\[
F_{\infty}(\x)=\exp\{\frac{i}{2\pi}\int_{\gam_d\cup \bar \gam_d}(s-e_1)\frac{\log{h(s)}}{\om_{+}(s)}ds\}
\]
with
\be \label{e1}
e_1=\frac{E+\bar E+d+\bar d}{2}.
\ee
\par
To account for this singularity, let us introduce the normalized, that is, its $a-$period vanishes, Abelian integral $w(\lam)$ of the second kind with simple poles at $\infty_{\pm}$:
\[
w(\lam)=\int_E^\lam \frac{z^2-e_1z+e_0}{\om_(z)}dz,
\]
where $e_1$ is the same as in (\ref{e1}) and $e_0$ is determined by the condition $\int_adw(\lam)=0$:
\[
e_0=-\frac{\int_d^{\bar d}(z^2-e_1z+e_0)\frac{dz}{\om_(z)}}{\int_d^{\bar d}\frac{dz}{\om_(z)}}.
\]
The large-$\lam$ expansion of $w(\lam)$ is of the form
\[
w(\lam)=\lam+w_{\infty}(\x)+O(\lam^{-1}),\qquad \lam\rightarrow \infty,
\]
where
\be \label{ominfty}
\begin{split}
w_{\infty}&=\int_E^{\infty}[\frac{z^2-e_1z+e_0}{\om_(z)}-1]dz-E\\
&=\frac{1}{2}(\int_E^{\infty}+\int_{\bar E}^{\infty})[\frac{z^2-e_1z+e_0}{\om_(z)}-1]dz-B
\end{split}
\ee
The jump conditions for $w(\lam)$ are as follows:
\[
\begin{split}
w_+(\lam)+w_-(\lam)=0,&\qquad \lam\in \gam_d\cup \bar \gam_d,\\
w_+(\lam)-w_-(\lam)=B_w,&\qquad \lam\in \gam_{\mu}.
\end{split}
\]
Here $B_w$ is the $b-$period of $w(\lam)$:
\be \label{Bom}
B_w=\int_bdw=2\int_E^d\frac{z^2-e_1z+e_0}{\om_(z)}dz=(\int_E^d+\int_{\bar E}^{\bar d})\frac{z^2-e_1z+e_0}{\om_(z)}dz\in \R.
\ee
Now introduce
\be \label{ellhatF}
\hat F(\lam)=F(\lam)e^{-i\Dta w(\lam)},
\ee
This new function is clearly bounded at $\lam=\infty$:
\be \label{ellFinfty}
\hat F(\infty,\x)=e^{i\hat \phi(\x)}.
\ee
with
\[
\hat \phi(\x)=\frac{1}{2\pi}\int_{\gam_d\cup \bar \gam_d}(s-e_1)\log{[h(s)\dta^{-2}(s,\x)]}\frac{ds}{\om_{+}(s)}-\Dta(\x)w_{\infty}(\x).
\]
Also, $\hat F(\lam)$ has the same jumps as $F(\lam)$ across $\gam_d$ and $\bar \gam_d$. On the other hand, the price for introducing the exponential factor in (\ref{ellhatF}) is that $\hat F(\lam)$ has a jump across $\gam_{\mu}$:
\[
\frac{\hat F_+(\lam)}{\hat F_-(\lam)}=e^{-i\Dta B_w},\qquad \lam\in \gam_{\mu}.
\]
Now we can absorb $\hat F(\lam)$ into the  Riemann-Hilbert problem for $N^{(4)}(x,t,\lam)$:
\[
N^{(4)}(x,t,\lam)=\hat F^{\sig_3}(\infty)N^{(3)}(x,t,\lam)\hat F^{-\sig_3}(\lam),
\]
which leads to the jump conditions
\[
N^{(4)}_+(x,t,\lam)=N^{(4)}_-(x,t,\lam)J_N^{(4)}(x,t,\lam),
\]
where
\[
J_N^{(4)}(x,t,\lam)=\left\{\ba{ll}J_N^{mod}+O(e^{-\varplon t}),&\quad \lam\in \gam_d\cup \bar \gam_d \cup \gam_{\mu},\\\id+O(e^{-\varplon t}),&\quad \lam\in L\cup \bar L.\ea
               \right.
\]
with
\be \label{ellJmod}
J_N^{(mod)}=\left\{\ba{ll} \left(\ba{cc}0&1\\1&0\ea \right),&\quad \lam\in \gam_d\cup \bar \gam_d,\\
                         \left(\ba{cc}e^{-itB_{\hat g}-i\Dta B_w}&0\\0&e^{itB_{\hat g}+i\Dta B_w}\ea \right),&\quad \lam\in \gam_{\mu},
                 \ea
          \right.
\ee

\subsubsection{The model problem}

Thus, we arrive at the model Riemann-Hilbert problem:
\begin{subequations}\label{ellMmodRHP}
\be \label{ellMmod}
N^{mod}_+(x,t,\lam)=N^{mod}_-(x,t,\lam)J_N^{mod}(x,t,\lam),\qquad \lam\in \gam_d\cup \bar \gam_d \cup \gam_{\mu},
\ee
\be \label{ellMmodasy}
N^{mod}(x,t,\lam)=\id+O(\frac{1}{\lam}),\qquad \lam\rightarrow \infty.
\ee
\end{subequations}
The solution of this model Riemann-Hilbert problem approximates $N^{(4)}(x,t,\lam)$:
\be \label{ellM4asyMmod}
N^{(4)}(x,t,\lam)=(\id+O(t^{-\frac{1}{2}}))N^{mod}(x,t,\lam),
\ee
The model problem (\ref{ellMmodRHP}) can be solved in terms of elliptic theta functions. Let
\[
U(\lam)=\frac{1}{c}\int_E^\lam\frac{dz}{\om(z)}
\]
be the normalized Abelian integral, that is
\[
c=2\int_{\bar d}^d\frac{dz}{\om(z)}
\]
Then, define
\be \label{tau}
\tau=\tau(\x)=\frac{2}{c}\int_E^d\frac{dz}{\om(z)}
\ee
with $\im \tau>0$. Furthermore, the following relations are valid:
\be \label{ellUkrel}
\begin{split}
\ba{ll}
U_+(\lam)+U_-(\lam)=0,&\quad \lam\in \gam_d,\\
U_+(\lam)+U_-(\lam)=-1,&\quad \lam\in \bar \gam_d,\\
U_+(\lam)-U_-(\lam)=\tau,&\quad \lam\in \gam_{\mu},
\ea
\end{split}
\ee
Next, define
\[
\nu(\lam)=(\frac{(\lam-E)(\lam-d)}{(\lam-\bar E)(\lam-\bar d)})^{\frac{1}{4}},
\]
where the branch is fixed by specifying the branch cut $\gam_{E,\bar E}$ and the behavior as $\lam\rightarrow \infty$;
\[
\nu(\lam)=1+\frac{D+d_2}{2i\lam}+O(\lam^{-2}),\qquad \lam\rightarrow \infty.
\]
Along the cut,we have
\[
\nu_+(\lam)=\left\{\ba{ll}-i\nu_-(\lam),&\quad \lam\in \gam_d\cup \bar \gam_d,\\-\nu_-(\lam),&\quad \lam\in \gam_{\mu}.\ea \right.
\]
Finally, introduce the theta function
\[
\tha_3(z)=\sum_{m\in \Z}e^{\pi i \tau m^2+2\pi i m z},
\]
and define the $2\times 2$ matrix-value function $\Theta(\lam)=\Theta(t,\x,\lam)$ with entries:
\[
\ba{c}
\Theta_{11}(\lam)=\frac{1}{2}[\nu(\lam)+\frac{1}{\nu(\lam)}]\frac{\tha_3[U(\lam)-U_0-\frac{1}{2}-\frac{B_{\hat g}t}{2\pi}-\frac{B_w\Dta}{2\pi}]}{\tha_3[U(\lam)-U_0]},\\
\Theta_{12}(\lam)=\frac{1}{2}[\nu(\lam)-\frac{1}{\nu(\lam)}]\frac{\tha_3[U(\lam)+U_0+\frac{1}{2}+\frac{B_{\hat g}t}{2\pi}+\frac{B_w\Dta}{2\pi}]}{\tha_3[U(\lam)+U_0]},\\
\Theta_{21}(\lam)=\frac{1}{2}[\nu(\lam)-\frac{1}{\nu(\lam)}]\frac{\tha_3[U(\lam)+U_0-\frac{1}{2}-\frac{B_{\hat g}t}{2\pi}-\frac{B_w\Dta}{2\pi}]}{\tha_3[U(\lam)+U_0]},\\
\Theta_{22}(\lam)=\frac{1}{2}[\nu(\lam)+\frac{1}{\nu(\lam)}]\frac{\tha_3[U(\lam)-U_0+\frac{1}{2}+\frac{B_{\hat g}t}{2\pi}+\frac{B_w\Dta}{2\pi}]}{\tha_3[U(\lam)-U_0]},
\ea
\]
where $U_0$ is to be chosen so that the unique zero of $\tha_3(U(\lam)-U_0)$, as a function on the Riemann surface, lying on the first sheet is compensated by the zero of $\nu(\lam)+\frac{1}{\nu(\lam)}$ where $\tha_3(U(\lam)+U_0)$ has no zero on this sheet. Setting
\[
U_0=U(E_0)+\frac{1}{2}+\frac{\tau}{2},
\]
where
\[
E_0=\frac{Ed-\bar E \bar d}{E-\bar E+d-\bar d}
\]
satisfies this requirement, and thus $\Theta(\lam)$ can be viewed as a function analytic in $\C\backslash \gam_{E,\bar E}$. On the other hand, due to the properties of theta function:
\[
\tha_3(-z)=\tha_3(z),\quad \tha_3(z+1)=\tha_3(z),\quad \tha_3(z\pm \tau)=e^{-\pi i \tau\mp 2\pi i z}\tha_3(z)
\]
$\Theta(\lam)$ satisfies the jump conditions (\ref{ellMmod})-(\ref{ellJmod}) of the model Riemann-Hilbert problem. Taking into account the normalization condition (\ref{ellMmodasy}), the solution of the model Riemann-Hilbert problem is given by
\[
N^{mod}(x,t,\lam)=\Theta^{-1}(t,\x,\infty)\Theta(t,\x,\lam).
\]

\subsubsection{Back to the original problem}

Now, following the sequence of equations of type (\ref{planebackq}) (with $g$ and $F$ replaced, respectively, by $\hat g$ and $\hat F$) and taking into account the equations $\hat g$ and $\hat F$, and the explicit formula for $n^{mod}_{12}(x,t,\lam)$
\[
2in^{mod}_{12}(x,t,\lam)=[D+d_2]\frac{\tha_3[\frac{B_{\hat g}t}{2\pi}+\frac{B_w\Dta}{2\pi}+U_0+\frac{1}{2}+U(\infty)]}{\tha_3[\frac{B_{\hat g}t}{2\pi}+\frac{B_w\Dta}{2\pi}+U_0+\frac{1}{2}-U(\infty)]}\frac{\tha_3[U_0-U(\infty)]}{\tha_3[U_0+U(\infty)]}
\]
and $\hat F^{-2}(\infty)=e^{-2i\hat \phi(\x)}$, we obtain the asymptotics in the region $-4t(B+\sqrt{2A^2(B+\frac{A^2}{4})})<x<-4tB$.

\begin{theorem}({\bf Elliptic wave region})
In the region $-4t(B+\sqrt{2A^2(B+\frac{A^2}{4})})<x<-4tB$,the asymptotics, as $t\rightarrow +\infty$, of the solution $q(x,t)$ of the initial value problem (\ref{DNLSandInit}) takes the form of a modulated elliptic wave:
\be \label{planefinalq}
q(x,t)=[D+\im d(\x)]\frac{\tha_3[\frac{B_{\hat g}t}{2\pi}+\frac{B_w\Dta}{2\pi}+V_+(\x)]}{\tha_3[\frac{B_{\hat g}t}{2\pi}+\frac{B_w\Dta}{2\pi}+V_-(\x)]}\frac{\tha_3[V_-(\x)-\frac{1}{2}]}{\tha_3[V_+(\x)-\frac{1}{2}]}+O(t^{-\frac{1}{2}}), t\rightarrow +\infty.
\ee

Here $B_{\hat g},B_w$ and $\Dta$ are functions of the variable $\x=\frac{x}{4t}$ defined, respectively, by (\ref{Bhatg}), (\ref{Bom}) and (\ref{Dta}), and $V_{\pm}(\x)=U_0+\frac{1}{2}\pm U(\infty)$. Furthermore,
\[
\tha_3(z)=\sum_{z\in \Z}e^{\pi i \tau m^2+2\pi i m z}
\]
is the theta function of invariant $\tau=\tau(\x)$ defined in (\ref{tau}),
\[
\hat g(\infty,\x)=t(2(\int_{E}^{\infty}+\int_{\bar E}^{\infty})[(z-\mu(\x))\sqrt{\frac{(z-d(\x))(z-\bar d(\x))}{(z-E)(z-\bar E)}}-(z+\x)]dz+2D^2-2B^2-4B\x)
\]
and the phase shift $\phi(\x)$ is given by
\[
\phi(\x)=\frac{1}{2\pi}\int_{\gam_{d}\cup \gam_{\bar d}}\frac{[s-e_1(\x)-\om_{\infty}(\x)]\log [h(s)\sqrt{s}\dta^{-2}(s,\x)]}{[(s-E)(s-\bar E)(s-d(\x))(s-\bar d(\x))]^{1/2}}ds
\]
where
\begin{equation*}
\ba{l}
h(\lam)=\left\{\ba{ll}a_+^{-1}(\lam)a_-^{-1}(\lam),&\lam\in\gam_d\\a_+(\lam)a_-(\lam),&\lam\in\gam_{\bar d}\ea \right.\\
\dta(\lam,\x)=\exp\{\frac{1}{2\pi i}\int_{-\infty}^{\mu(\x)}\frac{\log(1+\lam\rho^2(\lam))}{s-\lam}ds\}.
\ea
\end{equation*}
and $e_1(\x),\om_{\infty}$ and $\mu(\x)$ are defined, respectively, by (\ref{e1}), (\ref{ominfty}) and (\ref{mud2intBsys}).

\end{theorem}

\par
{\bf Acknowledgments.}
J.X want to thank Prof. V.P.Kotlyarov for the helpful suggestions.

\end{document}